\documentclass[twocolumn]{aastex631}
\usepackage{amsmath}

\usepackage{CJK}         
\usepackage{comment}
\usepackage{array}

\newcommand{\pfrac}[2]{\left( \frac{#1}{#2} \right)}
\newcommand{\ergss}{\rm erg \, s^{-1}}
\newcommand{\kms}{\rm km \, s^{-1}}
\newcommand{\mspyr}{\rm M_\odot \, yr^{-1}}

\newcommand{\rt}{r_{\rm t}}
\newcommand{\MBH}{M_{\rm BH}}

\begin{document}
\begin{CJK*}{UTF8}{gbsn}

\title{Star-Disk Collisions: Implications for Quasi-periodic Eruptions and Other Transients Near Supermassive Black Holes}

\author[0000-0003-3024-7218]{Philippe Z. Yao}
\affil{Department of Astrophysical Sciences, Princeton University, Peyton Hall, Princeton, NJ 08544, USA}

\author[0000-0001-9185-5044]{Eliot Quataert}
\affil{Department of Astrophysical Sciences, Princeton University, Peyton Hall, Princeton, NJ 08544, USA}

\author[0000-0002-2624-3399]{Yan-Fei Jiang (姜燕飞)}
\affil{Center for Computational Astrophysics, Flatiron Institute, New York, NY 10010, USA}

\author[0000-0002-1568-7461]{Wenbin Lu}
\affil{Department of Astronomy and Theoretical Astrophysics Center, University of California, Berkeley, CA 94720, USA}

\author{Christopher J. White}
\affil{Center for Computational Astrophysics, Flatiron Institute, New York, NY 10010, USA}
\affil{Department of Astrophysical Sciences, Princeton University, Peyton Hall, Princeton, NJ 08544, USA}

\begin{abstract}

We use \texttt{Athena++} to study the hydrodynamics of repeated star-accretion disk collisions close to supermassive black holes, and discuss their implications for the origin of quasi-periodic eruptions (QPEs) and other repeating nuclear transients.   We quantify the impact of the collisions on the stellar structure, the amount of stripped stellar debris, and the debris' orbital properties. We provide simple fitting functions for the stellar mass-loss per collision; the mass-loss is much larger after repeated collisions due to the dilute stellar atmosphere shock-heated in earlier collisions.   The lifetime of the QPE-emitting phase set by stellar mass-loss in star-disk collision models for QPEs is thus at most $\sim 1000$ years; it is shortest for eRO-QPE2, of order a few decades.   The mass of the stripped stellar debris per collision and its orbital properties imply that currently observed QPEs are not powered by direct star-disk collisions but rather by collisions between the stellar debris liberated in previous collisions and the accretion disk (`circularization shocks').  We discuss how the hydrodynamics of this interaction can explain the diverse timing properties of QPEs including the regular timing of GSN 069 and eRO-QPE2 and the large flare-to-flare timing variations observed in eRO-QPE1.   QPEs with recurrence times of many days, if observed, may have more regular timing.

\end{abstract}


\section{Introduction} \label{sec:intro}

\begin{figure*}
    \centering
    \includegraphics[width=0.75\textwidth]{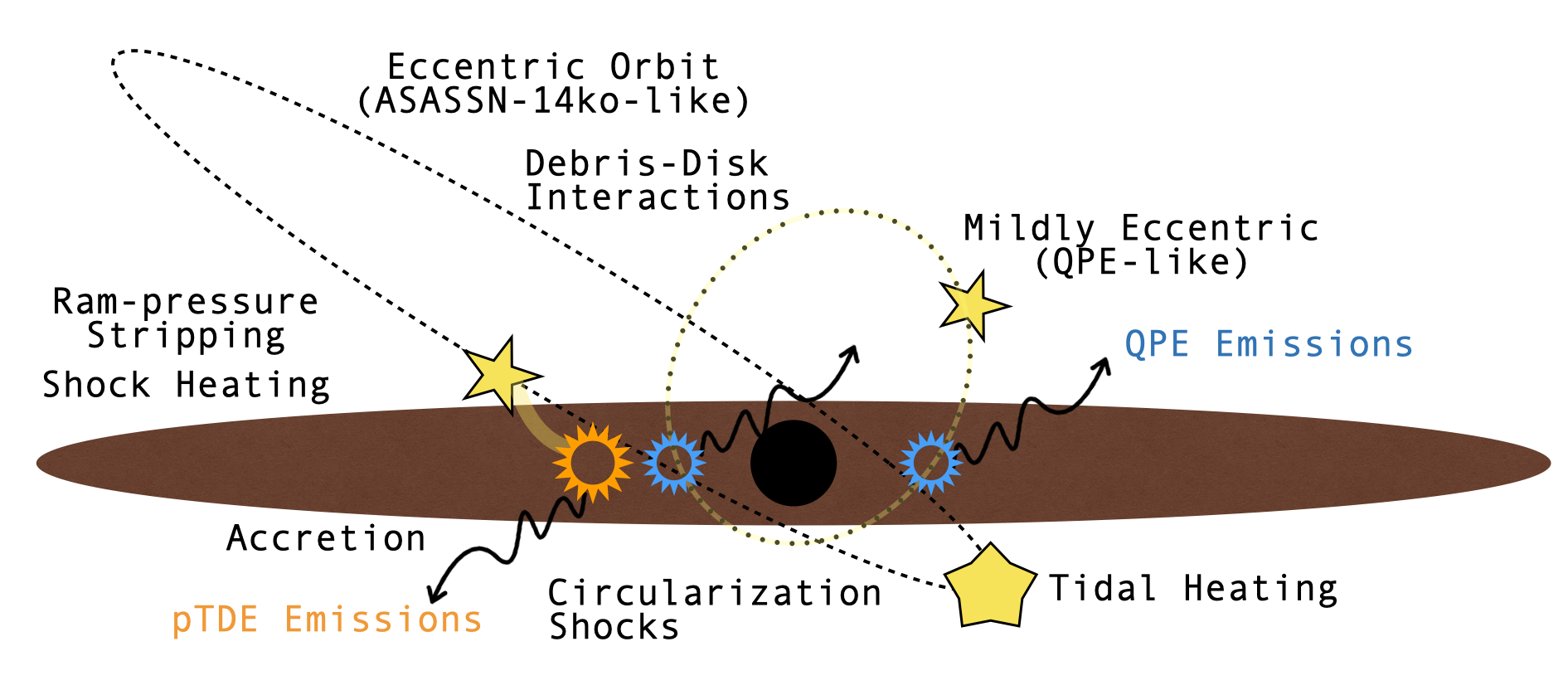}
    \caption{Schematic diagram showing physical ingredients in models of repeating nuclear transients involving a stellar companion to a SMBH. Sources of quasi-periodic eruptions (QPEs) may be stars on orbits with modest eccentricity (dotted line).   Partial tidal disruption (pTDE) candidates such as ASASSN-14ko (dashed line) likely have more eccentric orbits and much longer orbital periods.  Note that objects and orbits are not to scale.   In this paper we quantify the mass stripped from a star in star-disk collisions, the unbound stellar debris's orbital properties, and its possible contribution to repeating nuclear transients.}
    \label{fig:schematic}
\end{figure*}

The properties of stars orbiting close to supermassive black holes (SMBHs) can be dramatically modified by their proximity to the BH and by interaction with a gaseous accretion disk.   Tidal heating of the star can change its internal structure leading to rapid expansion of the star and mass transfer earlier than would otherwise be anticipated (e.g., \citealt{Li2013}).   If a dense geometrically thin accretion disk is present around the BH, star-disk collisions can dissipate orbital energy and angular momentum, leading to stellar inspiral much faster than via gravitational wave emission alone \citep{Syer1991,MacLeod2020,Linial2024}.  The same star-disk collisions strip mass from the star, potentially fueling the disk with which it interacts and destroying the star in the process \citep{Lu2023,Linial2023,Linial2024b}.

The discovery of nearly a dozen quasi-periodic large-amplitude recurring flares from galactic nuclei over the last $\sim 5$ years has generated renewed interest in the fate of stars on tightly bound orbits close to SMBHs.   These repeating nuclear transients (RNTs) include quasi-periodic X-ray eruptions (QPEs), repeating soft X-ray flares with recurrence times of hrs-days, rich and varied timing behavior, and no bright optical/UV counterparts \citep[e.g.][]{Miniutti2019, Arcodia2021, Chakraborty2021, Chakraborty2024, Arcodia2022, Miniutti2023, Webbe2023, Arcodia2024, Nicholl2024}; they occur in the nuclei of galaxies hosting low-mass SMBHs  \citep[$\sim 10^5 - 10^6 M_{\odot}$, e.g.][]{Wevers2022}.  In parallel, optical surveys have discovered longer-duration recurring nuclear transients, often classified as partial tidal disruption events (pTDEs). ASASSN-14ko is a particularly spectacular example \citep{Payne2021}, with $\sim$20 outbursts recorded to date and flares in the optical, UV, and X-rays \citep{Payne2022}. The flares last for $\sim$10 days, have a radiated energy of $\sim 10^{50}$ ergs, and have a mean period of $P_0 = 114.2\pm 0.4$ days, which decreases with a $\dot P \approx -0.0026 \approx -(7{\rm h})/P_0$ \citep{Payne2022}, much faster than what gravitational wave inspiral of a companion star can explain \citep{Payne2021, Linial2024}. 

Many features of repeating nuclear transients are suggestive of a binary origin, involving a star (or perhaps solar-mass compact object) orbiting a SMBH (see, e.g., \citealt{Xian2021, Sukova2021,Krolik2022,Zhao2022, Linial2023,Linial2024b,Lu2023,Tagawa2023, Franchini2023}).  While accretion disk instabilities might explain some features of RNTs \citep{Miniutti2019, Sniegowska2020, Raj2021, Pan2022, Kaur2023}, in this paper we continue to explore the hypothesis that RNTs involve a stellar companion to the SMBH, likely interacting with an accretion disk (there is clear evidence for quasi-steady disk emission in between the flares of many of the RNTs). The accretion disk can be fed from the star itself \citep{Lu2023}, be produced by a separate TDE \citep{Linial2023}, or be an otherwise unrelated AGN disk \citep{Sukova2021,Xian2021}.   
The radiation in RNTs is produced by some combination of (see the schematic diagram in Fig. \ref{fig:schematic}) (1) disk material shocked twice per orbit by the orbiting star, (2) circularization shocks as stellar debris stripped by star-disk interactions or the tidal gravity of the SMBH interacts with the disk, (3) accretion of such stripped material onto the BH.    It appears likely that the emission mechanisms are different for shorter-period QPEs and longer-period pTDE candidates.   For example, the short periods of QPEs are naively incompatible with the long viscous time required for accretion (\citealt{Lu2023}; see, however, \citealt{Raj2021}), while the large energies of the flares in ASASSN-14ko are incompatible with shocked disk material in star-disk interactions powering the emission \citep{Linial2024} (a promising model for QPEs; \citealt{Linial2023}).

In this paper, we quantify the properties of mass stripped from a star in star-disk collisions. We do so using hydrodynamical simulations with \texttt{Athena++}. For simplicity, we neglect the gravity of the BH in our calculations, which thus apply only to orbits that are at least a factor of few outside the tidal radius.  The calculations we present are important for understanding the role of star-disk collisions in producing repeating nuclear transients; we elaborate on this application throughout the paper, particularly the application to models of QPEs.    Our calculations also bear on the fate of all stars near SMBHs with dense gaseous disks.   Because our focus in this paper is on the hydrodynamics of the star and the stellar debris during star-disk collisions, our results do not apply to models in which QPEs are powered by compact object-accretion disk collisions.

The calculations we perform are similar to those used to study mass loss from the surviving companions of supernovae in binary systems \citep[e.g.][]{Liu2015, Bauer2019}.   The primary difference is that the ram pressure associated with star-disk interactions is generally lower than those produced by supernovae in close binary systems.  The star also undergoes repeated shocks twice per orbit so its structure is not that of a standard star.  Finally, the energy and angular momentum distribution of the unbound stellar debris is particularly important, as this determines how the stellar debris will interact with the disk later in its orbit, potentially powering the flares that we observe as repeating nuclear transients.

The remainder of this paper is organized as follows. We outline our simulation methods in \S \ref{sec:sim} and report the results of the simulations in \S \ref{sec:results}, including how star-disk interactions modify the structure of the star, cause stellar mass loss, and shock both the disk and unbound stellar debris to high energy. In \S \ref{sec:dE}, we adopt an analytic approach to further model aspects of the post-collision stellar debris dynamics. In particular, we discuss the implications of our results for QPE models in \S \ref{sec:QPE}, including their predicted lifetime and how stellar debris-disk interactions can contribute to the observed diversity of QPE timing behavior. We discuss the limitations of our simulations and summarize our findings in \S  \ref{sec:summary}.

\section{Simulation Methods} \label{sec:sim}

We carried out 2D and 3D simulations using the \texttt{Athena++} ideal hydrodynamics code \citep{Stone2020} in spherical polar coordinates with a statically refined logarithmic grid. The calculations are carried out in the frame of the star, centered on a $1M_{\odot}$, $1 R_\odot$ star that is initially a $\gamma = 5/3$ polytrope (we discuss the scalings to other mass stars below). The propagation of the shock into the star halts where the local stellar pressure $\sim p_{\rm ram}$.  For a $\gamma = 5/3$ polytrope the depth $\delta \ell$ varies with pressure as $\delta \ell/R_* \simeq (p/p_\star)^{2/5}$ where $p_\star = G M_\star^2/(4 \pi R_\star^4) \simeq 10^{15} \, {\rm erg \, cm^{-3}}$ is the mean stellar pressure.  The ram pressures we consider are within a factor of $10$ of $p_{\rm ram} \sim 10^{12} \, {\rm erg \, cm^{-3}}$, motivated by the ram pressures experienced by stars in the QPE models of \citet{Linial2023,Lu2023}.  The shocks will thus not reach deeper than a depth of $\delta \ell/R_\star \sim 0.05$.  To be conservative, we place the inner boundary at $0.9R_{\odot}$ (test simulations with the inner boundary at $0.8 R_\odot$ yield nearly identical results). We find that the structure of the star is indeed unmodified by the shocks near our inner boundary (see, e.g., Fig. \ref{fig:1DProfile} discussed below).   Given that most of the mass is a point mass, rather than being on the domain, we include gravity via a spherically symmetric point mass potential of fixed stellar mass.   We adopt an equation of state (EOS) that includes both gas- and radiation-pressure using \texttt{Athena++}'s general EOS module \citep{Coleman2020}; this is important because some of the shocked material is dominated by radiation pressure, not gas pressure. 

In addition to the star, we initialize a disk with a Gaussian density profile and scale height $H_{\rm disk}$; we define the latter as half of the Gaussian's full width half maximum, consistent with the usual accretion disk definition of $H_{\rm disk}$ as the half-thickness of the disk.  Note that for radiation pressure dominated disks $H_{\rm disk} \simeq 3 R_\odot (M_{\rm BH}/10^6 M_\odot) (\dot M/0.1 \dot M_{\rm Edd})$ where $\dot M_{\rm Edd} = L_{\rm Edd}/0.1 c^2$.  We consider values of $H_{\rm disk} \simeq 0.175-5.6 R_\odot$, as well as a few simulations at even larger $H_{\rm disk}$.    The peak disk density $\rho_{\rm disk}$ is initially centered at a fixed $z-$value in our spherical coordinate system; the center of the disk is a distance $R_{\rm disk}$ away from the star along the polar axis. The disk is then initialized with a velocity along the $z-$axis towards the star at $\sim 0.03-0.1 c$. As we describe below, the region close to the stellar surface ($\lesssim 2 R_{\odot}$) is most refined to resolve the surface layer of the stellar envelope where most dynamic processes take place; further refinement is extended to large distances from the star ($\lesssim 20 R_{\odot}$) to minimize numerical diffusion of the disk as it approaches the star.  
All disks have an initially constant sound speed $c_{\rm s, disk} = 0.1 v_{\star}$ throughout, where $v_\star$ is the speed of the disk towards the star (in the star's reference frame).  This ensures that the shock propagating into the disk from the star-disk collision is supersonic.

To better distinguish between disk and stellar material as they collide and mix, we initialize two unique passive scalars associated with each component. This allows us to isolate the stellar and disk material in the Figures that follow, subject to the usual numerical mixing inherent to finite-volume numerical schemes. 

Star-disk collisions differ from the analogous problem of supernovae ejecta impacting a binary companion in that the structure of the star is modified by the cumulative impact of many collisions.   As a result, we carry out multiple star-disk collisions until the properties of the newly stripped stellar material reach an approximate steady state.   Since QPEs recur on a timescale of a few hours, we set the collision interval to 5 hours for most of our simulations.  The new disk created for each subsequent collision has the same properties and is positioned far away from the star at $R_{\rm disk} = 14 R_{\odot}$, in order to minimize its impact on the unbound stellar debris created in previous collisions (we discuss the limitations of this approach in \S \ref{sec:dE}).   We carried out tests with a factor of two longer intervals between collisions and initializing the incoming disk at somewhat different distances from the star.  Our results are not very sensitive to the specific choices used here (see Fig \ref{fig:convergence}).   In our simulations, the disk impacts the same side of the star in each simulation.  In reality, the part of the star colliding with the disk will vary. Our simplified collision geometry is reasonable, however, because after $\sim 3-4$ collisions the shocked star's structure is relatively spherically symmetric (Fig. \ref{fig:athsims}) and so the results of subsequent collisions are independent of the exact star-disk collision orientation.

Our 2D simulations are initialized with a ($128 \times 64$) root grid in $(r, \theta)$, with $r_{\rm inner} = 0.9 R_{\odot}$ and $r_{\rm outer} = 1000 R_{\odot}$. We then refine the inner $0.9 - 1.5\,R_{\odot}$ by a factor of $32\times$ in each dimension.  The level of refinement varies roughly logarithmically in radius until reaching the root grid resolution outside $\sim 24\,R_{\odot}$. Our fiducial resolution corresponds to $dr/r \simeq  1.7  \times 10^{-3}$ in the most highly refined region, i.e., $\sim 20$ cells across the initially shocked surface layers of the star.   After several shocks the surface layer of the star heats up leading to a larger scale-height; our resolution is effectively better for these later simulations. We also carry a convergence test with individually higher levels of refinement, greater  $R_{\rm disk}$ to further minimize overwriting stripped stellar debris, and a set of 3D simulations. We will discuss the results of the convergence tests in \S \ref{sec:stellarmassloss}.

Our 3D simulations are initialized with a ($64 \times 32 \times 64$) root grid in $(r, \theta, \phi)$, with $r_{\rm inner} = 0.9 R_{\odot}$ and $r_{\rm outer} = 100 R_{\odot}$, and the inner $0.9 - 1.5\,R_{\odot}$ refined by a factor of $8\times$, corresponding to a highest resolution region with $dr/R \simeq 10^{-2}$. This is sufficient to resolve the scale height of the stellar atmosphere after several shocks have increased the scale height but the resolution is poor for the first 3D star-disk interaction. Between $1.5 - 12\,R_{\odot}$, the root grid is refined by a factor of $4\times$ to prevent the disk from diffusing within the possible collision area. Note that the grid along the polar direction is also refined by the corresponding amount within each defined radial boundary. 

\subsection{Scaling to Other Stellar Masses}
\label{sec:othermass}

The key dimensionless parameters in our problem can be expressed as (1) $M_\star/M_{\rm disk}$ where $M_{\rm disk} = 2 \pi R_\star^2 H_{\rm disk} \rho_{\rm disk}$ is the local disk mass intercepted by the star, (2) $H_{\rm disk}/R_\star$, and (3) $v_\star/v_{\rm esc,\star}$, where $v_{\rm esc,\star}$ is the stellar escape speed.    We thus expect that for stars of different mass and radii, the results of simulations with the same values of these dimensionless parameters should be similar (assuming the star is still initially modeled as a $\gamma = 5/3$ polytrope).   This scale-invariance is not rigorously true, however, because of our equation of state (gas and radiation pressure).    As a check on whether the simulations presented here can also be applied to other stellar masses, we ran a simulation of a $0.3 M_\odot$, $R_\star = 0.38 R_\odot$ ($R_\star \propto M_\star^{0.8}$ for the low-mass main sequence) star with $v_\star = 0.088 c$, $H_{\rm disk} = 0.53 R_\odot$ and $\rho_{\rm disk} = 5.4 \times 10^{-6} {\rm g \, cm^{-3}}$.   This is a scaled lower stellar mass version of our fiducial simulation.   The mass-loss per collision was within about $50 \%$ of the analogous $1 M_\odot$ star collision.  In what follows we present our $1 M_\odot$ star results in physical units appropriate for that stellar mass but give fitting functions for an arbitrary $M_\star$.

\subsection{Thermal Readjustment of the Star}
\label{sec:thermal}

The simulations presented in this paper do not include radiation transfer.   The motivation for neglecting radiation is that the dynamical interaction between the star and the disk takes place on less than or of order a stellar dynamical time, much shorter than the thermal time of even the outer layers of the star.   It is thus reasonable to neglect radiation during the relatively prompt star-disk interaction that we focus on in this paper.

Nonetheless, radiative losses after the star-disk interaction are potentially important if the thermal time of the shocked surface layers of the star is much shorter than the orbital period.  In that case, the appropriate stellar structure to consider for the star is similar to that of a typical star (because it loses thermal memory of being shocked between collisions).  By contrast, if the thermal time of the shocked surface layers is longer than the orbital period, then the appropriate stellar structure to consider is one strongly modified by the cumulative effects of many star-disk interactions.    To estimate the appropriate regime, we note that the thermal time at optical depth $\tau$ is roughly $t_{\rm th} \simeq (p_{\rm tot}/p_{\rm rad}) \tau H/c$ where $H$ is the scale-height at optical depth $\tau$ and $p_{\rm tot}$ and $p_{\rm rad}$ are the total (gas + radiation) and radiation pressure, respectively.   The optical depth and pressure in the envelope are in turn related by $\tau \simeq p \kappa/g$ where we have assumed that the stellar surface layers of interest are physically thin (thickness $\lesssim R_*$) and $\kappa$ and $g$ are the typical opacity and surface gravity, respectively.   We are interested in the thermal time at the place where the stellar pressure is of order the ram pressure of the shock produced by the star-disk interaction (perhaps somewhat lower pressure if the shock is on for less than a stellar dynamical time).   The thermal time of the shocked envelope $t_{\rm th}$ relative to the stellar dynamical time $t_{\rm dyn}$ can thus be estimated as
\begin{equation}
    \begin{aligned}
    \frac{t_{\rm th}}{t_{\rm dyn}} & \sim \left( \frac{p_{\rm ram} t_{\rm dyn} \kappa}{c} \right) \left( \frac{H}{R_*} \right) \left( \frac{p_{\rm tot}}{p_{\rm rad}} \right) \\
    & \sim 5 \times 10^4 \left( \frac{p_{\rm ram}}{10^{12} \, {\rm erg \, cm^{-3}}} \right) \left( \frac{t_{\rm dyn}}{1 \, {\rm hr}} \right) \left( \frac{H}{R_*} \right) \left( \frac{p_{\rm tot}}{p_{\rm rad}} \right)
    \label{eq:thermal}
    \end{aligned}
\end{equation}
In the second line above we have assumed electron scattering opacity and scaled to a sun-like star and typical ram pressures in the QPE models of \citet{Linial2023,Lu2023}; the model for gas-drag induced orbital decay in ASASSN-14ko requires $p_{\rm ram}$ larger by several orders of magnitude \citep{Linial2024}. The last factor in equation \ref{eq:thermal} is $\gtrsim 1$.   The surface layers are also shocked to $H \sim R_*$.  Equation \ref{eq:thermal} thus implies that for orbital periods $P \ll$ yr the shocked surface layers have $t_{\rm th} \gg P$ and so the appropriate stellar structure to consider is one strongly modified by repeated shocks.  This applies to stars on short orbital periods around SMBHs, as in QPE models.   By contrast, for longer orbital periods as expected in the optically selected pTDE candidates, the outer layers of the star can thermally readjust some between collisions and so an unperturbed stellar structure may be a better approximation (depending on the exact $p_{\rm ram}$ and orbital period).   We will consider both limits in this work to bracket the range of star-disk interactions expected near SMBHs.

\section{Results} \label{sec:results}

In what follows we describe the properties of the shocked star and unbound stellar debris produced in repeated star-disk collisions.   As described in \S \ref{sec:thermal}, the impact of repeated star-disk interactions in the vicinity of SMBHs depends on the orbital period of the star.   For short orbital periods (e.g., QPEs with orbital periods of hours to days), there is no time for the outer layers of the star to thermally adjust to the repeated shocks it encounters.   To model this regime we carry out a series of multiple star-disk collisions until the properties of the stripped mass in each encounter reach a rough statistical steady state.   On the other hand, for longer orbital periods (e.g., pTDEs with periods of years), the star can radiate away some or all of the thermal energy produced in a collision by the time it next encounters the disk.   The stellar structure in each collision is then more similar to that of a normal (unshocked) star.   In our simulations, this regime is best modeled by our first collision between the star and the disk.   

The unbound stellar debris created in each star-disk collision expands away from the star over time.   In reality, such gas will start to feel the SMBH's gravity once it exits the star's Hill sphere.   We do not have this dynamics in our simulations since we only include the star's gravity.   In \S \ref{sec:dE} we use the results of our simulations to analytically estimate the evolution of the stellar debris including the effects of the BH's gravity. 

Table \ref{tab:params} summarizes the most important physical parameters in our set of simulations presented here. 

\renewcommand\arraystretch{1.2}
\begin{table}
    \centering
        \begin{tabular}{ |l|l|l|l| }
         \hline
         \multicolumn{4}{|c|}{Model Parameters} \\
         \hline
         \hline
        $v_\star\,(c)$ & $\rho_{\rm disk}\,(\rm g/cm^3)$ & $H_{\rm disk}\,(R_{\odot})$ & $t_{\rm cross}/t_{sh}$\\
         \hline
            0.03 - 0.1 & $10^{-6}$ - $10^{-5}$ & 0.175 - 5.6 & 0.01 - 0.5\\
         \hline
        \end{tabular}
        \caption{Range of model parameters we explore in this paper. The fiducial simulation has $v_\star\ = 0.1c$,  $\rho_{\rm disk} = 10^{-6} \rm g/cm^3$, $H_{\rm disk} = 1.4R_{\odot}$, and $t_{\rm cross}/t_{\rm sh} \simeq 0.1$.   $t_{\rm cross}/t_{\rm sh}$ measures the time for the star to cross the disk relative to the time it takes the shock to stall, i.e., reach pressure balance, in the stellar interior (eq. \ref{eq:tshratio}).}
        \label{tab:params}
\end{table}

\begin{figure*}
    \centering
    \includegraphics[width=0.32\textwidth]{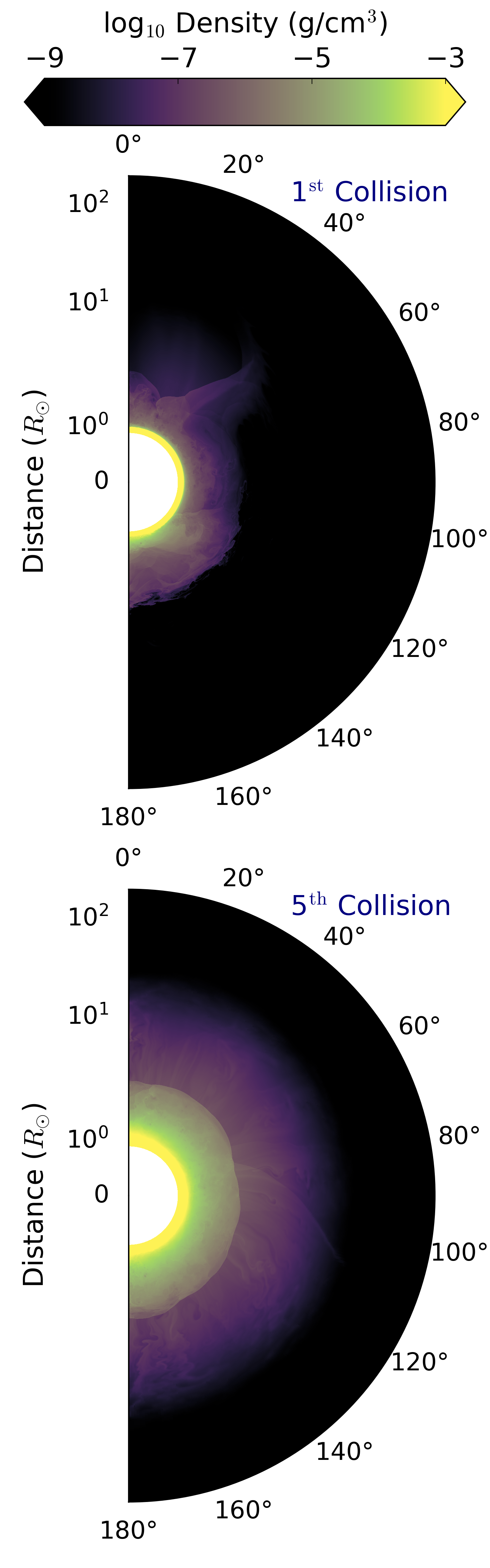}
    \includegraphics[width=0.32\textwidth]{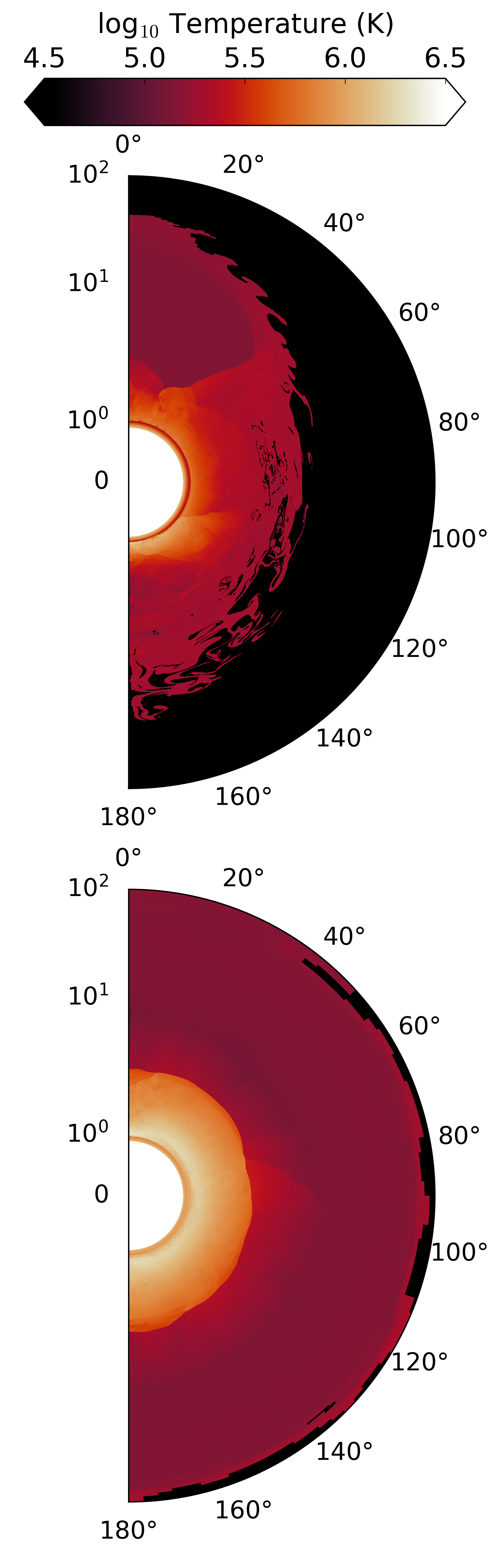}
    \includegraphics[width=0.32\textwidth]{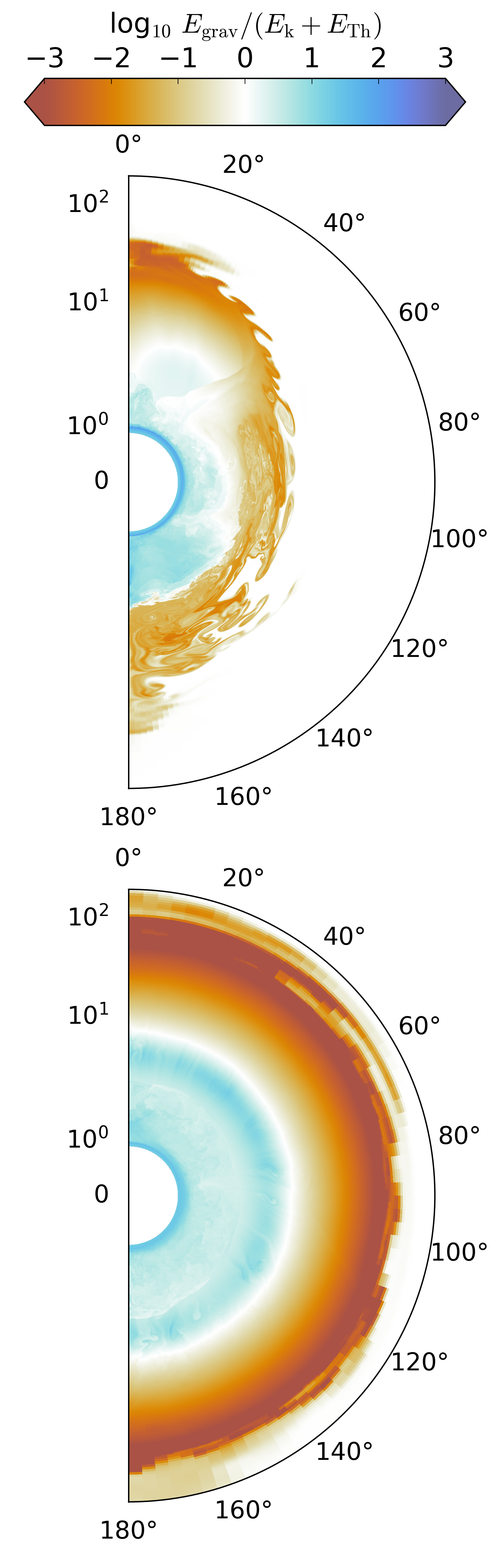}  
    \caption{Density, temperature, and gravitational binding energy distributions (\textit{left} to \textit{right}) after a star has been shocked 1 (\textit{top}) and 5 (\textit{bottom}) times by an ambient disk in our fiducial 2D \texttt{Athena++} simulations refined at level 7 ($128\times$). After a few collisions, the star is surrounded by a nearly spherical extended envelope still bound to the star (blue in the energy plot) and an expanding cloud of unbound stellar debris (orange/brown).}    
    \label{fig:athsims}
\end{figure*}

\begin{figure*}
    \centering
    \includegraphics[width=\textwidth]{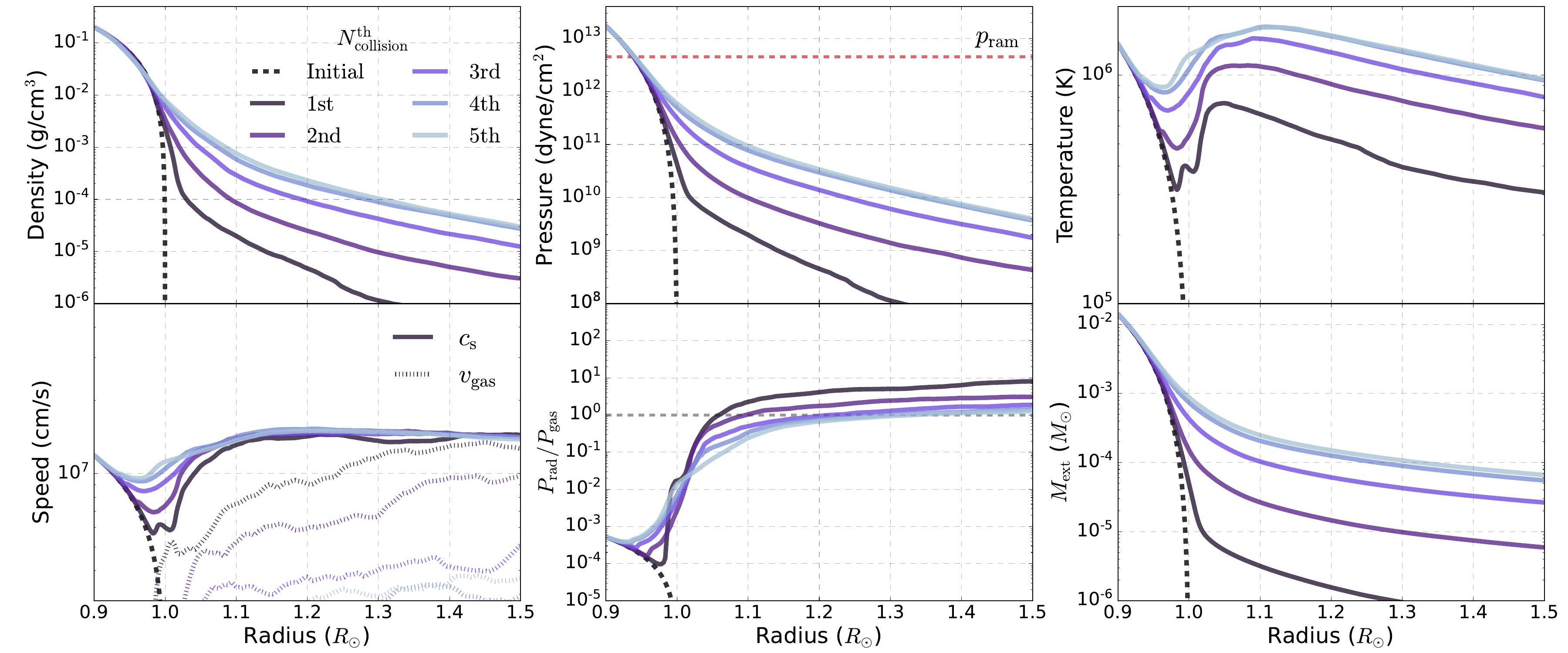}
    \caption{Angle-averaged density, pressure, temperature, sound speed, radiation to gas pressure profiles, and integrated exterior bound mass as a function of stellar radius for our fiducial simulation after the star experiences 1 to 5 disk collisions (also shown in Fig \ref{fig:athsims}). Lighter colors represent an increasing number of collisions. The red dashed line in the pressure panel shows the characteristic disk ram pressure in this simulation ($p_{\rm ram}$). The speed panel compares the local gas velocity $v_{\rm gas}$ (dotted) to the local sound speed $c_{\rm s}$ (solid). The initial polytropic stellar profile is shown in black dashed lines.  After $\sim 3-4$ collisions the bound outer envelope of the star that has been repeatedly shocked converges to a roughly constant sound speed profile. The interior of the star remains unscathed.}
    \label{fig:1DProfile}
\end{figure*}

\begin{figure*}
    \centering
    \includegraphics[width=\textwidth]{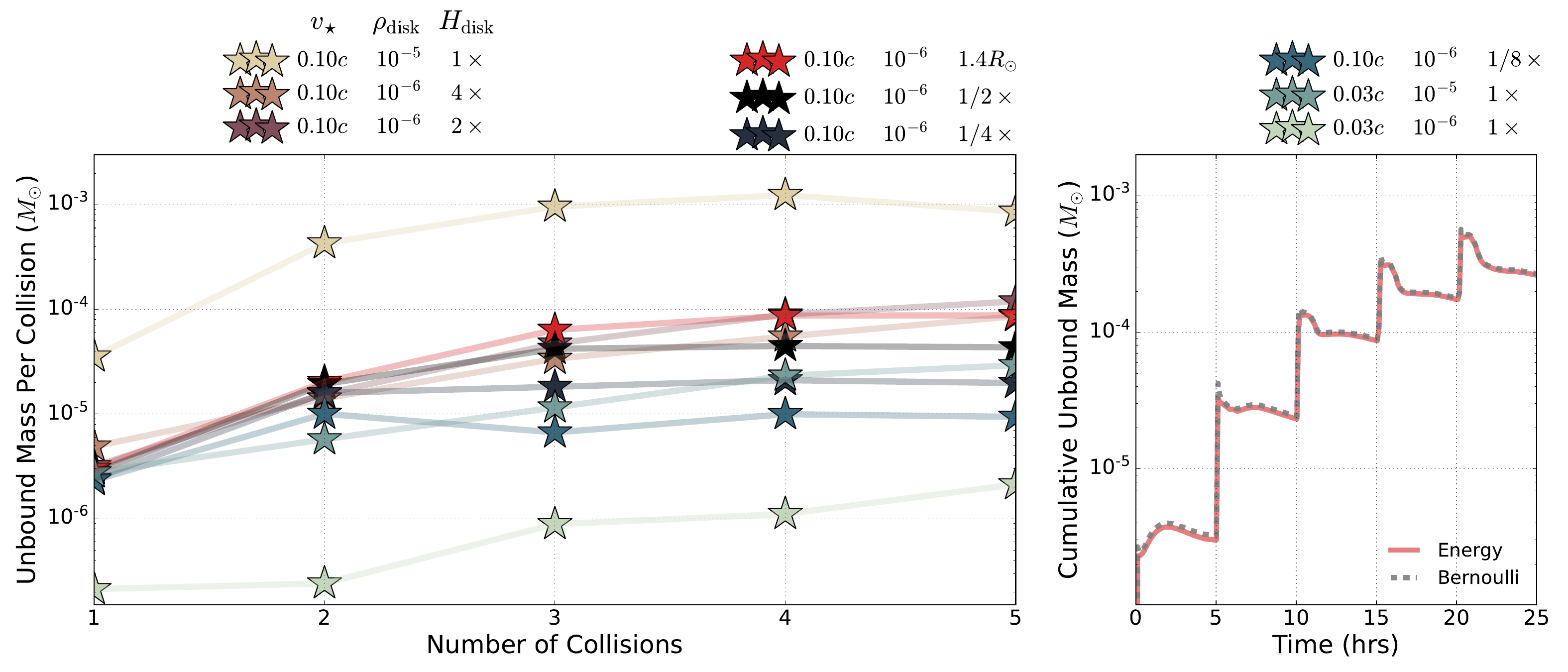}
    \caption{\textit{left:} Mass stripped from the star per disk-collision for a range of physical parameters: $v_{\star}$ is the orbital velocity of the star at the point of collision, $\rho_{\rm disk}$ is the central gas density of the disk (in g/cm$^3$), $H_{\rm disk}$ is the disk thickness relative to the value for the fiducial simulation (solid red), for which $H_{\rm disk} = 1.4 R_{\odot}$.  The unbound stellar mass per collision increases significantly for the second and third collisions since the extended shocked stellar envelope is easier to strip than the initial polytropic profile; by $\sim 4-5$ collisions the unbound mass per collision is relatively independent of the number of additional collisions.  Fits to these results are given in Equation \ref{eq:1stfit} and \ref{eq:fit}. \textit{right:} Cumulative instantaneous unbound stellar mass as a function of time over 5 star-disk collisions for the fiducial simulation. The red and grey lines show the small impact of defining the unbound mass using specific energy or specific Bernoulli parameter.}    
    \label{fig:massloss}
\end{figure*}

\subsection{Shocked Stellar Structure} \label{sec:massloss}

\begin{figure}
    \centering
    \includegraphics[width=\columnwidth]{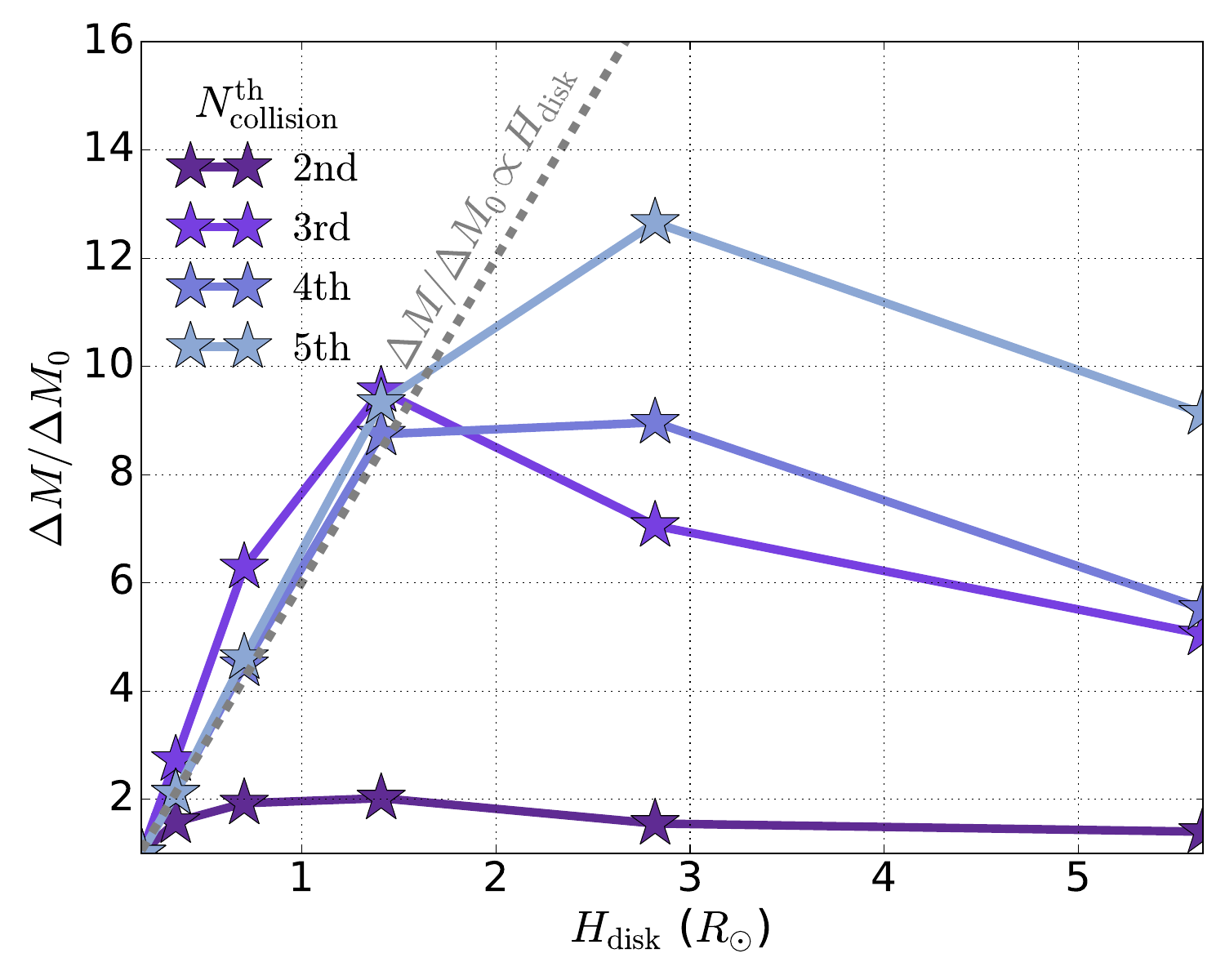}
    \caption{Dependence of the mass stripped per collision on the disk scale height for the 2D $\rho_{\rm disk} = 10^{-6} \, {\rm g cm^{-3}}$ and $v_{\star} = 0.1c$ simulations.  $\Delta M$ is normalized by $\Delta M_0$, the amount of mass loss after the $N^{\rm th}$ collision for the smallest $H_{\rm disk} = 0.175R_{\odot}$ we used.  After many collisions, the amount of mass loss is initially linear until saturating when the disk is sufficiently thick. These simulations do not accurately resolve the first collision, which is studied at higher resolution in Figure \ref{fig:1st_hit}.}   \label{fig:disk_height}
\end{figure}

\begin{figure}
    \centering
    \includegraphics[width=\columnwidth]{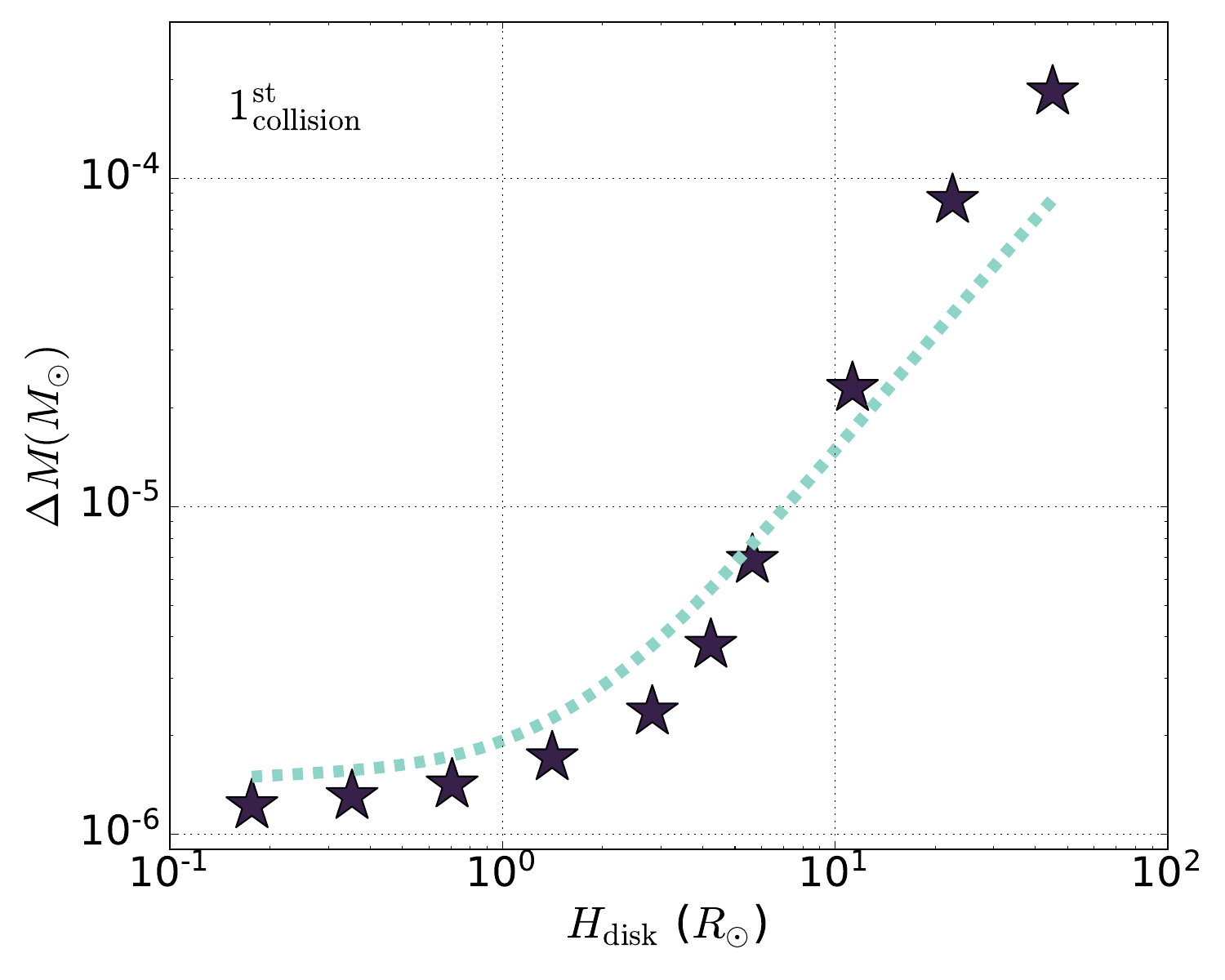}
    \caption{Dependence of the mass stripped per collision from an undisturbed star on the disk scale height for the 2D $\rho_{\rm disk} = 10^{-6} \, {\rm g cm^{-3}}$ and $v_{\star} = 0.1c$ simulations.  The numerical fit in equation \ref{eq:1stfit} is shown with a dotted line.}   \label{fig:1st_hit}
\end{figure}

We show density, temperature, and gravitational binding energy snapshots for our fiducial 2D simulations in Figure \ref{fig:athsims}. The 2 rows show the properties of the stellar material $\sim 5$ hrs after the 1st and 5th disk collision. The stellar material is isolated using a cut on the stellar passive scalar. The corresponding spherical shell-averaged 1D density, temperature, sound speed, pressure, radiation to gas pressure, and exterior mass profiles are also shown in Figure \ref{fig:1DProfile}. After the initial collision, the stellar envelope is quickly heated up by the ram pressure of the shock; the stellar envelope is significantly asymmetric after the first shock. However, after the fifth collision, both the bound and unbound stellar material are much more spherically distributed (this will happen even more quickly in reality due to the changing orientation between the star and disk over the course of multiple orbits).   The radius of the star, defined, e.g., as the transition from bound to unbound material, increases significantly as a result of the star-disk collisions, with an inner dense envelope now inflated to $\sim 2-3 R_\odot$. The cloud of unbound stellar debris also extends to much larger radii over time, as matter unbound in previous collisions continues to expand to larger radii. We note that our simulations adopt a density floor of $10^{-10}\, \rm g\,cm^{-3}$ and pressure floor of $10^6\, \rm dyne\,cm^{-2}$, which corresponds to a temperature of $\sim 10^5 \rm K$ in radiation dominated low-density regions. Hence, the measured temperature of shocked unbound stellar debris at large radii in Figure \ref{fig:athsims} is an upper limit assuming thermal equilibrium.

Figure \ref{fig:1DProfile} shows that the structure of the outer layers of the star is also dramatically modified by the repeated collisions, with the polytropic outer profile replaced by a roughly constant sound speed atmosphere having $c_s \sim 100 \, \kms$ and similar outflow velocities.  As a result, the amount of bound stellar mass in the outer layers of the star is larger than prior to the repeated collisions.   After the first collision much of the shocked envelope is radiation pressure dominated but as the envelope mass increases in subsequent collisions gas and radiation pressure are similar (radiation pressure dominates the lower density shocked material at larger radii not shown in Figure \ref{fig:1DProfile}).  Note also that the thermal structure of the envelope is closer to a convectively stable profile with $T \sim {\rm const}$ or $c_s \sim {\rm const}$ than to $T \propto \rho^{2/3}$ as would be expected if the solution had constant gas entropy with radius (shock heating imparts more entropy to the lower density outer layers, leading to a convectively stable profile). 

In addition to the mass lost by ablation and stripping during the star-disk collision, the inflated stellar envelope will also drive a Parker-like wind due to the large scale height of the stellar atmosphere.   Neglecting for the moment the surrounding unbound stellar debris, the physical conditions in Figure \ref{fig:1DProfile} would produce an isothermal stellar wind with a sonic point at $\sim 10 R_\odot$ and $\dot M \sim   10^{-5} \, \mspyr$.   As we show below, this is much less than the mass-loss rate produced directly during the star-disk collisions.   The high pressure and density in the unbound stellar debris would also likely suppress such a wind (i.e., the ambient bounding pressure cannot be neglected as in standard stellar wind theory).

The temperature map in Figure \ref{fig:athsims} shows an expanding sphere of hot stellar debris that extends to $\sim  100 R_{\odot}$ by the end of the fifth collision. This cloud of gas remains optically thick with temperatures at large radii $\sim 10^{5.5} \, \rm K$. This is too low to produce QPEs, which are soft X-ray flares with thermal-like spectra and temperatures of a few million K (e.g., \citealt{Miniutti2019}).    A more promising source of  QPEs is the disk material shocked during the star-disk collision \citep{Linial2023} or the disk material shocked by the unbound stellar debris as it returns and collides with the disk in later orbits; we return to this point in \S \ref{sec:dE}.

\subsection{Stellar Mass Loss} \label{sec:stellarmassloss}

The \textit{left} panel of Figure \ref{fig:massloss} shows the mass stripped from the star after each disk encounter for a variety of disk parameters and stellar velocities. We categorize a cell as unbound when its thermal and kinetic energy is sufficient to exceed the local gravitational potential.  The \textit{right} panel of Figure \ref{fig:massloss} shows the cumulative unbound mass for our fiducial simulation as a function of time, using two definitions of unbound mass (specific energy and specific Bernoulli parameter, respectively); the two definitions agree well, particularly after the very dynamic phase of the star-disk interaction (the `spike' in unbound mass).

Figure \ref{fig:massloss} shows that the unbound mass produced after each collision increases by a factor of $\sim 3-4$ between the first two collisions, and subsequently reaches a roughly constant value by about the 4th or 5th collision that is $\sim 20$ times larger than the mass-loss in the first collision (we carried out several of these runs for an additional 5 collisions and the mass loss per collision was similar to within a factor of $\lesssim 2$).   This increase is largely due to the fact that after multiple collisions the stellar envelope is significantly inflated with a much larger fraction of its mass at low densities (Fig. \ref{fig:1DProfile}) that mix more efficiently with the colliding disk material. Figure \ref{fig:massloss} shows that for the parameters simulated here, motivated by QPE models, between $10^{-5}-10^{-4} M_\odot$ per collision are unbound from the star.   We discuss the implications of this for the lifetime of stars producing QPEs in \S \ref{sec:lifetime} below.

Figure \ref{fig:disk_height} isolates the dependence of the mass loss per collision ($\Delta M$) on the thickness of the disk with which the star collides, for the 2nd-5th collisions.  $\Delta M$ is then normalized by $\Delta M_0$: the amount of mass loss for the smallest $H_{\rm disk}$ value. Figure \ref{fig:1st_hit} shows the same mass-loss per collision as a function of disk thickness, but for the first collision.  The first collision simulations here are carried out with 4 times higher resolution (7 levels of mesh refinement) to ensure that the thin surface layer of the star is adequately resolved. This higher resolution is not needed for the later collisions, as we show explicitly in Figure \ref{fig:convergence} below.

Figure \ref{fig:disk_height} shows that 
in the limit of multiple collisions the stripped mass scales $\propto H_{\rm disk}$ until $H_{\rm disk} \gtrsim R_\star$.  By contrast, Figure \ref{fig:1st_hit} shows that in the first collision, the dependence of the mass stripped on disk thickness is initially much weaker, but then eventually becomes roughly linear in $H_{\rm disk}$ for $H_{\rm disk} \gtrsim 3 R_\star$ (almost the opposite dependence of the multiple collision result!).

The amount of mass loss depends on the disk thickness because the disk thickness influences both the duration of the star-disk interaction and the total shock energy transferred from the disk to the star.   One way to quantify the interaction time is via the time for the star to cross the disk $t_{\rm cross} = 2H_{\rm disk}/v_\star$ compared to the time it takes the shock to reach a depth in the star where the pressure is similar to the shock ram pressure; for an $n = 3/2$ polytrope, the latter is (see Appendix \ref{sec:firstcoll} for more discussion).
\begin{equation}
t_{\rm sh} \simeq t_{\rm dyn} \left(\frac{p_{\rm ram}}{ p_\star}\right)^{1/5}
\label{eq:tsh}
\end{equation}
such that
\begin{equation}
\frac{t_{\rm cross}}{t_{\rm sh}} \sim \frac{R_\star/v_\star}{t_{\rm dyn}} \frac{2H_{\rm disk}}{R_\star} \left(\frac{p_{\rm ram}}{ p_\star}\right)^{-1/5}
\label{eq:tshratio}
\end{equation}
where $t_{\rm dyn} = \sqrt{R_\star^3/GM_\star}$ is the star's global dynamical time and $p_\star = GM_\star^2/4\pi R_\star^4$ is the mean stellar pressure. For our fiducial simulation with $H_{\rm disk} \sim R_\odot$ this yields $t_{\rm cross}/t_{\rm sh} \sim 0.1$.   In this regime of $t_{\rm cross}/t_{\rm sh} \lesssim 1$, one might expect the amount of mass stripped to depend on the disk thickness since this changes the depth that the shock can reach in the star.   

\begin{figure}
    \centering
    \includegraphics[width=\columnwidth]{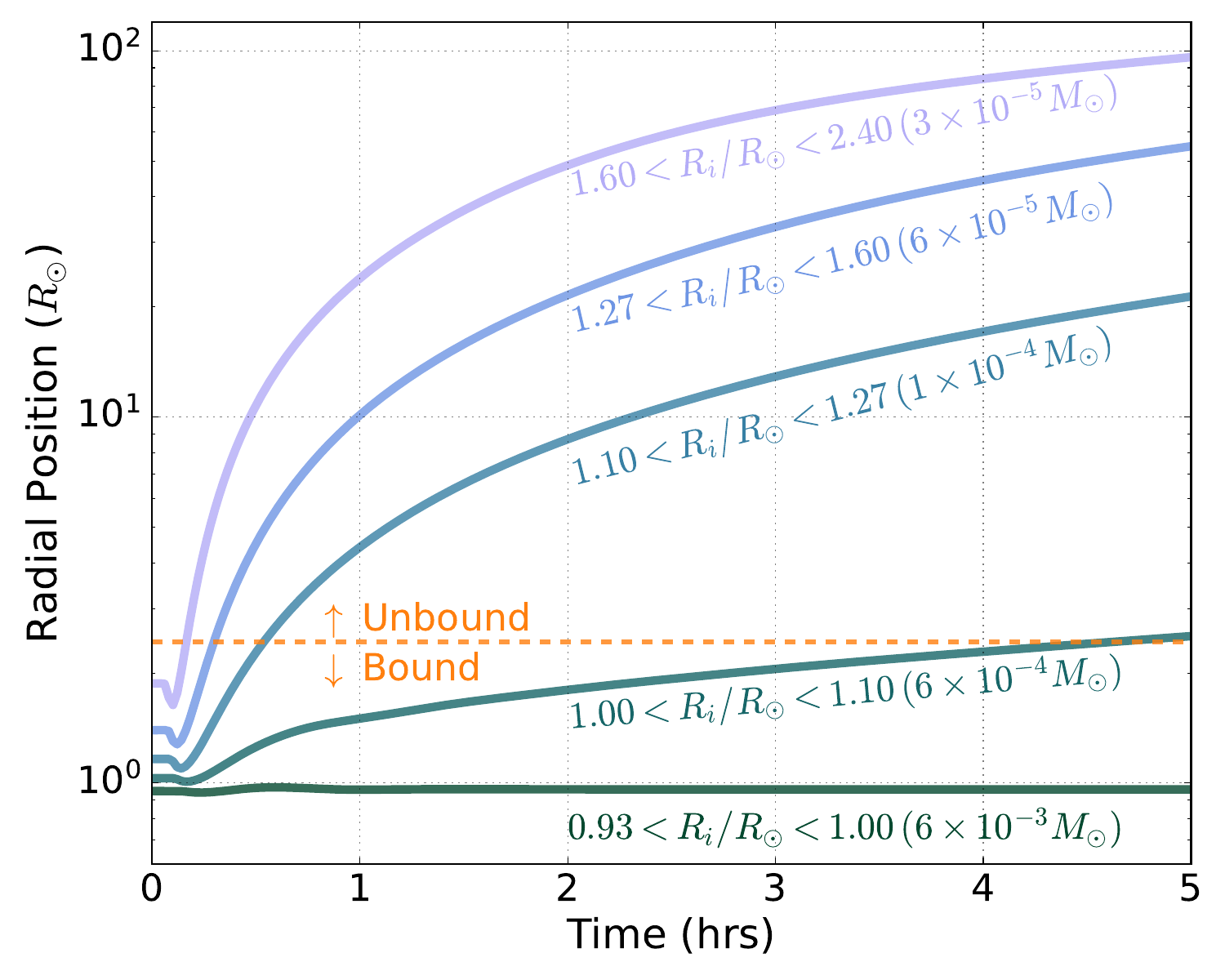}
    \caption{Evolution of the mass-weighted mean radial position of initially bound stellar layers throughout the 5th disk-collision in our fiducial simulation. $t=0$ is approximately 5 hours after the 4th disk collision. The dashed orange line shows the angle-averaged boundary between bound and unbound stellar material at $t=0$. Each layer is traced by a different passive scalar defined by the initial radial positions labeled in the legend (with the mass in each region also indicated). By the end of this disk collision, most of the outer initially bound layers ($\sim1.1 < R_i/R_{\odot} < 2.4$) are unbound, and the deeper bound stellar layers expand to replenish the stellar envelope, producing the roughly time-steady density and exterior mass profiles shown in Fig. \ref{fig:1DProfile}. }   \label{fig:masszones}
\end{figure}

Our interpretation of Figures \ref{fig:disk_height} and \ref{fig:1st_hit} is as follows.   We begin with the first collision results.   In the first collision, the stellar mass-loss depends weakly on disk thickness for small $H_{\rm disk}$ because the amount of the stellar envelope that is shocked by the star-disk interaction itself depends weakly on $H_{\rm disk}$ when $t_{\rm cross}/t_{\rm sh} \lesssim 1$.   We derive an approximate scaling in this regime in Appendix \ref{sec:firstcoll}.    Perhaps more interesting is the result for large disk thickness in which $\Delta M \propto H_{\rm disk}$.   This is consistent with a fixed (small) fraction of the total shock energy due to the star-disk interaction going into ablating mass from the star (the total disk mass swept up by the star scales $\propto H_{\rm disk}$ as does the total shock energy).  

After multiple collisions, the physics of the stellar mass loss is quite different because it is dominated by the disk interacting with the dilute stellar atmosphere created by the cumulative effect of previous star-disk interactions.  The linear dependence of stellar mass loss on $H_{\rm disk}$ for small $H_{\rm disk}$ in Figure \ref{fig:disk_height} again reflects a fixed fraction of the star-disk shock energy ablating the dilute stellar atmosphere.  The mass-loss saturates, i.e., no longer increases with increasing $H_{\rm disk}$, when most of the dilute stellar atmosphere is unbound in the star-disk interaction.  To demonstrate this explicitly, we ran the 5th disk collision of our fiducial simulation with additional passive scalars that trace different layers of the initially bound stellar material. The mass-weighted mean radial positions of these layers as a function of time are shown in Figure \ref{fig:masszones}.  Figure \ref{fig:masszones} shows that almost all of the previously bound stellar envelope at $R_i \gtrsim 1.1 R_\odot$ is unbound during the collision.   At the same time the deeper layers (e.g., $1 R_\odot \lesssim R_i \lesssim 1.1 R_\odot$) are inflated to larger radii, replenishing the bound envelope.\footnote{It is plausible that the mass-loss for multiple collisions will again increase with increasing $H_{\rm disk}$ for much larger disk scale-heights than shown in Figure \ref{fig:disk_height}; effectively once the dilute atmosphere is ablated the continued disk-star interaction in the limit of a very thick disk is analogous to the first collision and so should eventually scale as in Figure \ref{fig:1st_hit}.}    

Another consequence of Figure \ref{fig:masszones} is that, so long as the Hill sphere of the star is larger than $\sim 2-3 R_\star$ (where here $R_\star$ is explicitly the initial, unshocked, radius of the star), the tidal gravity of the BH is unlikely to be important for determining the mass unbound per collision (since gas exterior to $\sim 2-3 R_\star$ is unbound even absent the BH's tidal gravity).   If the Hill sphere is smaller than $\sim 2-3 R_\star$, however, the mass-loss per star-disk collision will be in a different regime not captured by the calculations in this paper; the BH's tidal gravity and tidal heating will be dynamically critical in changing the structure of the star (see the Appendix of \citealt{Linial2024c} for a detailed discussion of tidal heating in this context).

The simulation results in Figure \ref{fig:massloss}, \ref{fig:disk_height}, and \ref{fig:1st_hit} can be reasonably fit with two simple functions, one describing the results of the first collision and the other describing the result after many collisions:
\begin{equation}
\begin{aligned}
& \ \ \ \ \ \ t_{\rm th} \ll P \, ({\rm single \, shock):}\quad  \\
\frac{\Delta M_{\star}}{M_{\star}} &\simeq 0.001 \left(\frac{p_{\text{ram}}}{p_{\star}}\right) \Bigg[ \left( \frac{H_{\text{disk}}}{R_{\star}} \right)^{0.25} \\
&\quad - 1.2 \left( \frac{H_{\text{disk}}}{R_{\star}} \right)^{0.6} + 0.6 \left( \frac{H_{\text{disk}}}{R_{\star}} \right) \Bigg]
\end{aligned}
\label{eq:1stfit}
\end{equation}

\begin{equation}
\begin{aligned}
& \ \ \ \ \ t_{\rm th} \gg P \, ({\rm multiple \, shocks):}\quad  \\
\frac{\Delta M_{\star}}{M_{\star}} 
&\simeq 0.03 \left(\frac{p_{\rm ram}}{ p_\star} \right) \pfrac{H_{\rm disk}}{R_\star} \left[1 + \frac{H_{\rm disk}}{R_{\star}}\right]^{-1}\\
& \xrightarrow[H_{\rm disk} \lesssim R_\star]{} 0.03 \left(\frac{H_{\rm disk}}{R_{\star}} \right) \left(\frac{p_{\rm ram}}{ p_\star} \right) \\
& \xrightarrow[H_{\rm disk} \gtrsim R_\star]{} 0.03 \, \left(\frac{p_{\rm ram}}{ p_\star} \right) \\
 \frac{\Delta M_{\star}}{M_{\rm disk}} &\simeq 0.06 \, \frac{v^2_{\star}}{v^2_{\star,esc}} \left[1 + \frac{H_{\rm disk}}{R_\star}\right]^{-1}.
\label{eq:fit}
\end{aligned}
\end{equation}
We reiterate that the `first collision' regime is appropriate for longer period and/or lower ram pressure star-disk interactions where the star can thermally readjust between collisions and the `many collisions' regime is appropriate for shorter period and/or higher ram pressure star-disk interactions when the star cannot thermally adjust (see eq. \ref{eq:thermal}).   For the single shock fit in equation \ref{eq:1stfit} we prioritized matching the slopes of the simulation results at large and small $H_{\rm disk}$ so that modest extrapolation would be reasonable.  For the case of multiple collisions in equation \ref{eq:fit} we have given two equivalent expressions for $\Delta M_\star$, one relative to the stellar mass $M_\star$ and the other relative to the local disk mass that the star interacts with, $M_{\rm disk} = \pi R_\star^2 \Sigma_{\rm disk} = \pi R_\star^2 2 \rho_{\rm disk} H_{\rm disk}$.
The scatter of the simulated mass-loss rates relative to equation \ref{eq:fit} is roughly $50 \%$.    Note that the $H_{\rm disk} \gtrsim R_\star$ result in equation \ref{eq:fit} is a factor of 10 larger mass-loss per collision than assumed in previous work (e.g., \citealt{Linial2023,Linial2024}) based on \citet{Liu2015}'s supernovae-companion interaction simulations.  This highlights the large difference between shocking an undisturbed star versus one repeatedly shocked by multiple star-disk interactions.

\begin{figure}
    \centering
    \includegraphics[width=\columnwidth]{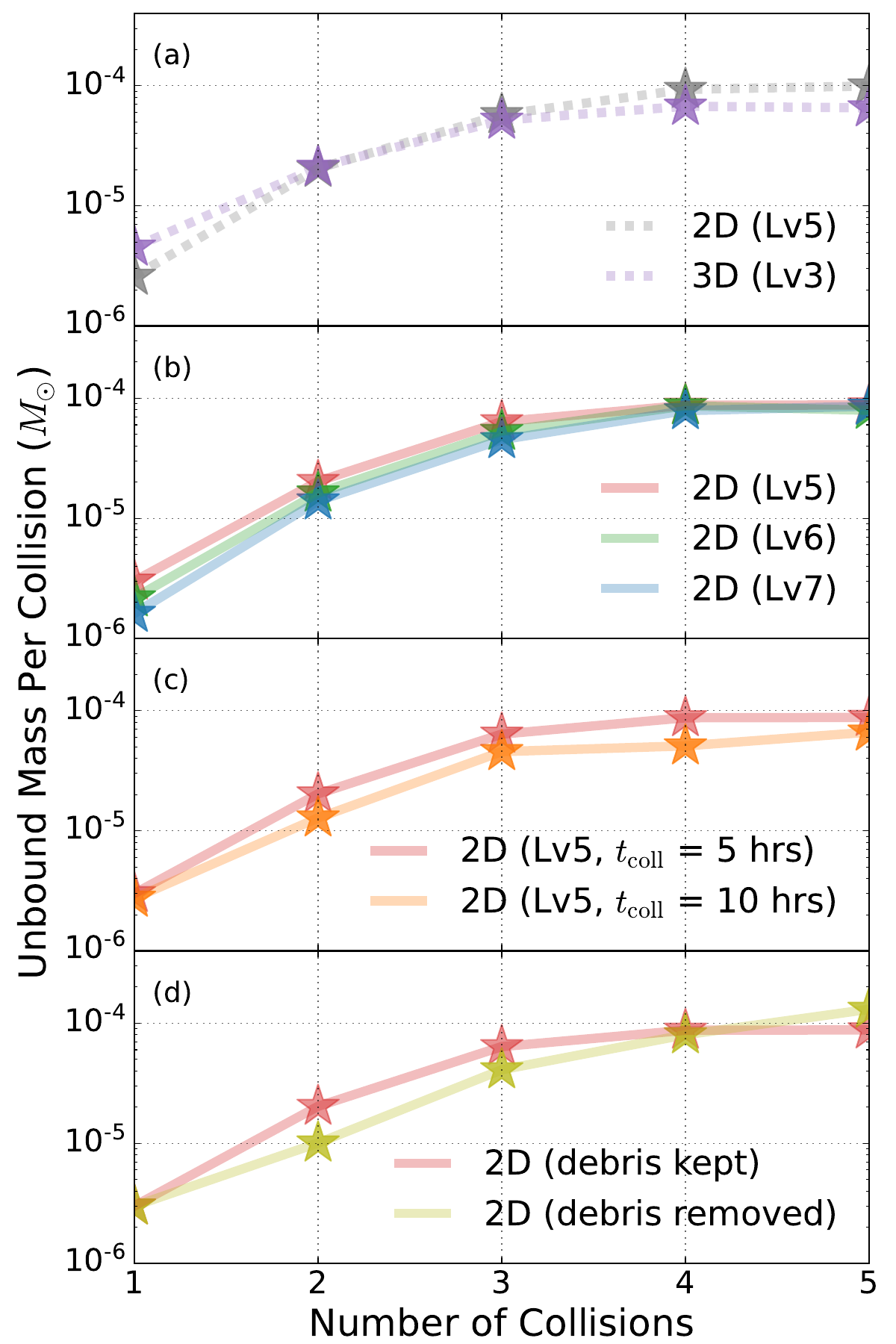}
    \caption{Convergence tests for unbound stellar mass per collision. Panel (a) compares a 3D simulation with 3 ($8\times$) levels of refinement to a 2D simulation with 5 ($32\times$) levels of refinement. Panel (b) compares 2D simulations with 5 ($32\times$), 6 ($64\times$), and 7 ($128\times$) levels of refinement. Panel (c) compares 2D simulations with 2 different collision intervals $t_{\rm coll}$ (5 and 10 hours). Panel (d) compares 2D simulations with and without unbound stellar debris artificially removed before the next disk encounter. All panels are for $\rho_{\rm disk} = 10^{-6} \, {\rm g cm^{-3}}$ and $v_{\star} = 0.1c$ simulations, with $H_{\rm disk} = 0.35 R_{\odot}$ and $t_{\rm coll} = 1 \,\rm hr$ in panel (a) and $H_{\rm disk} = 1.4 R_{\odot}$ and $t_{\rm coll} = 5 \,\rm hrs$ in panels (b)-(d).} 
    \label{fig:convergence}
\end{figure}

In Figure \ref{fig:convergence}, we test the convergence of our simulations by varying a few numerical parameters of our simulation with $v_{\star} = 0.1c$, $\rho_{\rm disk} = 10^{-6}$ g/cm$^3$. To minimize the time and refinement region of the computationally expensive 3D simulation, panel (a) uses a disk with $H_{\rm disk} = 0.35R_{\odot}$ and a collision interval $t_{\rm coll} = 1 \,\rm hr$, while in panels (b)-(d) $H_{\rm disk} = 1.4R_{\odot}$ and $t_{\rm coll} = 5 \,\rm hrs$. Panel (a) shows that 3D and 2D simulations give similar results, with the largest difference for the initial collision in which our 3D simulation does not resolve the thin stellar surface layer as well as the 2D simulation (so in this case, we believe that the 2D result is more reliable).  Panel (b) compares results for 5, 6, and 7 levels of static mesh refinement in our 2D simulations.  Higher resolution somewhat decreases the mass loss in the first star-disk collision, but the results are well-converged after many star-disk interactions when the stellar envelope is heated up and becomes more extended. Panel (c) compares two different time intervals between each disk collision, with results that are relatively insensitive to this change.  Panel (d) shows the effect of removing previously unbound stellar material before each subsequent collision. The change in the amount of mass loss is negligible.  We also carried out tests (not shown) varying the initial distance of the disk from the star (which changes how much unbound stellar debris the disk interacts with); this did not produce a significant change in the unbound mass per collision.  The bottom two panels of Figure \ref{fig:convergence} highlight that the amount of mass unbound from the star in later star-disk collisions is not sensitive to exactly how much of the previously unbound stellar debris each subsequent disk collision interacts with.    This result is important because the orbit of the unbound stellar debris is not necessarily modeled correctly in our simulations that neglect the BH's gravity. 

\begin{figure*}
    \centering
    \includegraphics[width=\textwidth]{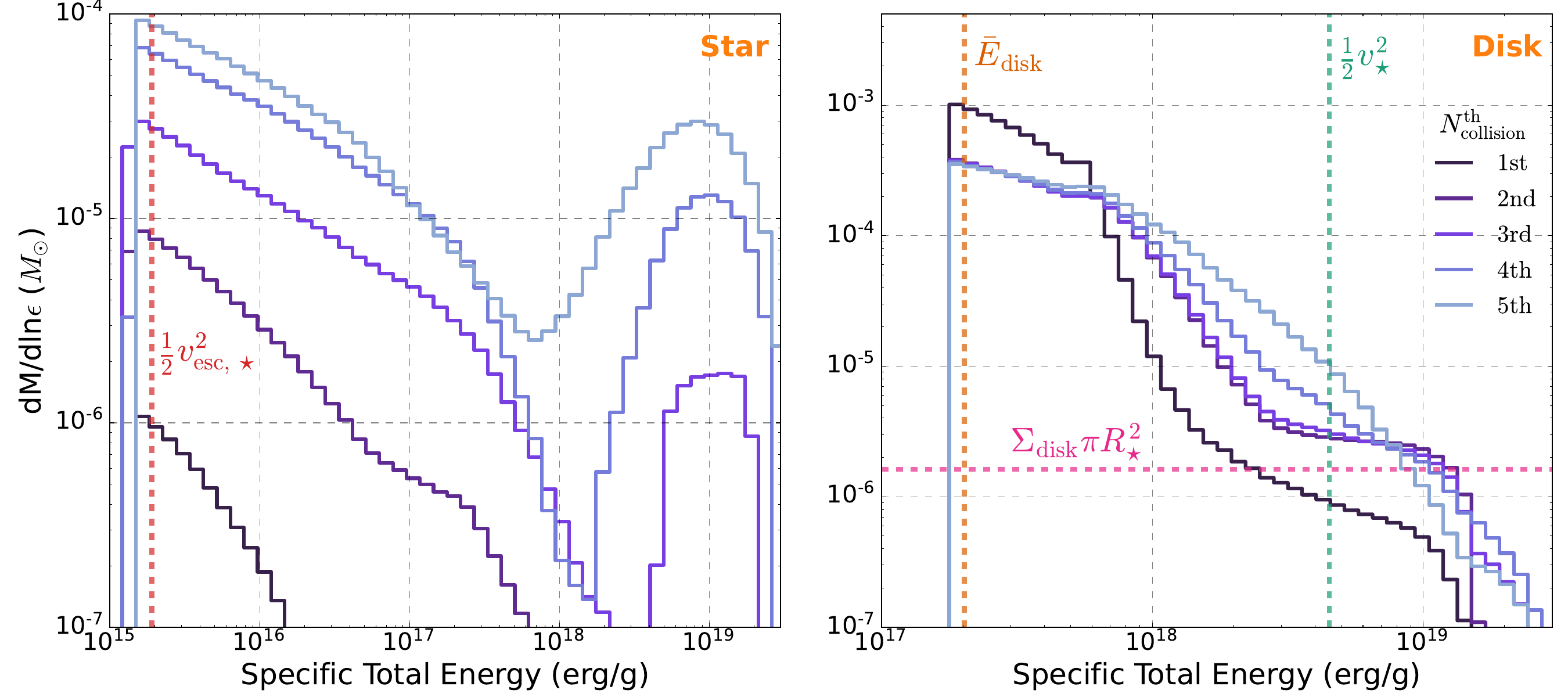}
    \caption{Energy distribution of unbound stellar debris (in the frame of the star) (\textit{left}) and disk (in the frame of the disk)  (\textit{right}) after 1-5 star-disk collisions.    Three key energy scales highlighted with vertical dashed lines are set by the stellar escape speed $v_{\rm esc,\star}$, the stellar orbital velocity $v_\star$, and the initial disk thermal energy $\bar E_{\rm disk}$. The unbound stellar debris has a typical energy set by $v_{\rm esc,\star}$; the high specific energy tail after multiple collisions is previous unbound debris shocked by the disk in later star-disk collisions (this high energy tail may not be quantitatively accurate because we do not include the BH gravity; see \S \ref{sec:dE}). In the right panel, undisturbed disk material remains at the initial disk thermal energy $\bar E_{\rm disk}$ while the shocked disk has a typical specific energy set by the stellar orbital velocity $v_\star$.   The amount of high energy shocked disk material significantly exceeds the naive estimate $\sim \Sigma_{\rm disk} \pi R_\star^2$ (horizontal purple line) in later collisions because the collision is dominated by the unbound stellar debris shocking with the disk, not the star itself; the former has a larger area than the star (see also Fig. \ref{fig:disk_symmetry}).}    
    \label{fig:energy_dist}
\end{figure*}

\subsection{Energetics of the Shocked Disk and Unbound Stellar Debris} \label{sec:energy}

\begin{figure*}
    \centering
    \includegraphics[width=\textwidth]{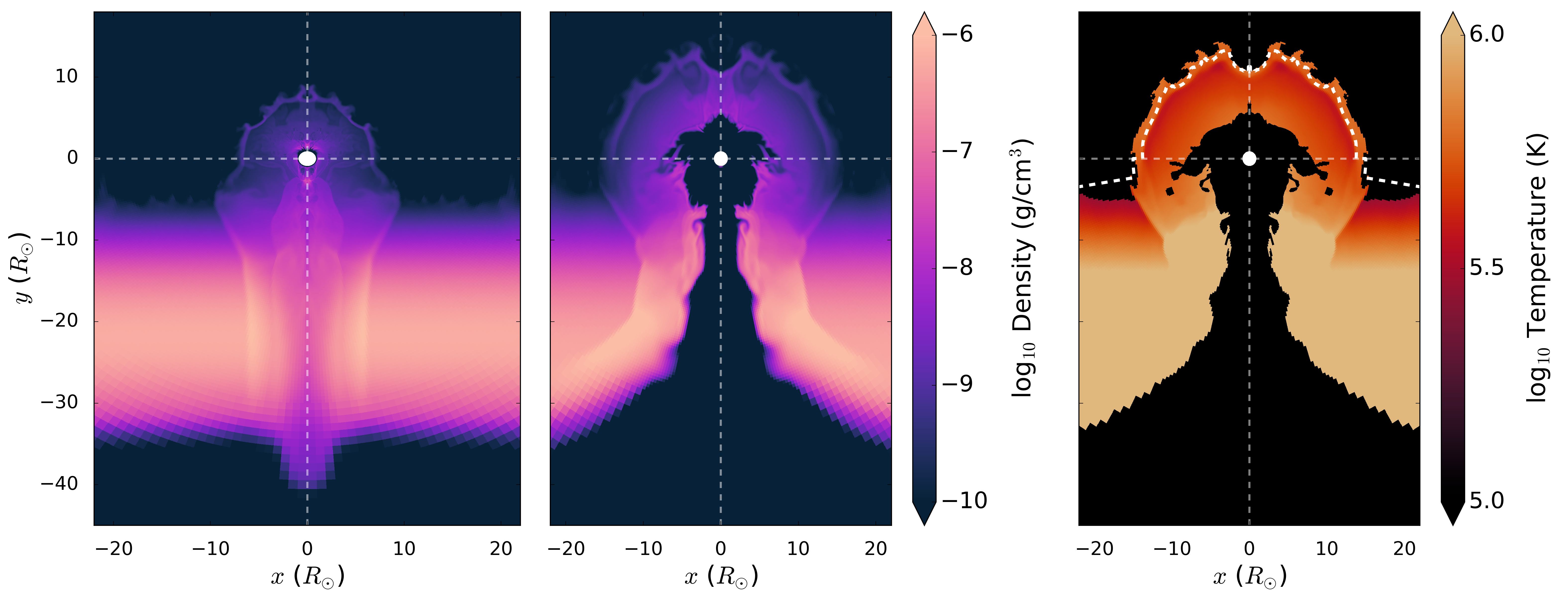}\\    
    \caption{Post-collision disk density map after encountering the star with no surrounding unbound debris ({\it left}) and the star with its unbound stellar debris from previous collisions ({\it center}). The temperature map corresponding to the latter case ({\it right}) is also shown with a dashed contour for the optical depth $\tau \sim c/v_\star \sim 10$ of the shocked disk. We show our disk model with $H_{\rm disk} = 2.8 R_{\odot}$ to highlight the qualitative difference in the shocked disk ejecta when $H_{\rm disk} > R_{\star}$ (\textit{left}) and $H_{\rm disk} \sim R_{\rm debris}$ (\textit{center}).  Only the former produces a somewhat symmetric ejection of disk material both in and opposite to the direction of the star's motion.   The temperature of the shocked disk near the photosphere $\sim 10^{5.7}$ K is too low to explain QPE flares in our models that assume thermal equilibrium between the gas and radiation.  Different levels of refinement in our spherical-polar grid cause the uneven grid resolution in this Cartesian plot.  The disk density is isolated in this plot from the stellar material using a passive scalar cut (this produces the evacuated region between the star and disk in the {\it center} and {\it right}  panels). }
    \label{fig:disk_symmetry}
\end{figure*}

In Figure \ref{fig:energy_dist}, we show the energy distribution of the unbound stellar and disk debris for a range of specific total energy, the sum of the local specific kinetic and thermal energy. The energy distributions are cumulative for the unbound stellar debris stripped during the 5 collisions we show here.  The stellar debris energy is measured in the frame of the star including only stellar matter that is not bound to the star (the star is isolated from the disk using a passive scalar cut). The disk debris energy is measured in the disk frame isolating the disk material using a relatively stringent cut on the disk passive scalar (this is necessary because the unbound stellar mass is much larger than the shocked disk mass so a modest amount of mixing contaminates diagnostics of the disk material).

For the shocked star, most of the debris has an energy of order $\frac{1}{2} v_{\rm esc,\star}^2$.  The puffy stellar envelope becomes easier to strip as it continues to encounter the disk, resulting in a significantly higher shocked mass over time.  In addition, the lower density of the previously shocked stellar debris implies that later debris-disk interactions create higher specific energy stellar debris.  This is why the high-energy component of the stellar debris increases over time in Figure \ref{fig:energy_dist}.   

A consequence of the expanding stellar debris is that in the 2nd and later star-disk collisions, the dominant effect on the disk is {\it not} the collision between the star and the disk, but rather the collision between the stellar debris from previous collisions and the disk.   We illustrate this in Figure \ref{fig:disk_symmetry} which shows the post-shock disk density for the first and fifth collisions in our simulations with $v_{\star} = 0.1c$, $\rho_{\rm disk} = 10^{-6}$ g/cm$^3$, and $H_{\rm disk} = 2.8R_{\odot}$ (the disk material is isolated from the stellar material using our disk and star passive scalars).   The shocked disk occupies a much larger area in the latter collision relative to the former.   A consequence of this is that the amount of disk mass at high specific energy increases in later collisions relative to the first collision, as shown explicitly in the right panel of Figure \ref{fig:energy_dist}: the mass of the shocked disk with specific energy $\frac{1}{2} v^2_{\star}$ after the first collision matches the expected intercepted disk mass $\sim \Sigma_{\rm disk} \pi R_{\star}^2 $ (purple dotted line).   
As more collisions lead to an expanding gas cloud of previously stripped stellar material around the star, the mass of the newly shocked disk continues to increase by about a factor of $\sim 10$, reaching $\sim 10^{-5} M_{\odot}$ for this choice of star and disk parameters.  We discuss the implications of this result for QPE models in the next section.   Note that the majority of the disk has a specific energy (in the disk frame) of $\bar E_{\rm disk}$ set by the thermal energy of the unshocked disk material.

Figure \ref{fig:disk_symmetry} ({\em right}) also shows the temperature of the expanding shocked disk material produced in our simulations that assume local thermal equilibrium between the gas and radiation.   The temperature is too low to explain QPE flares with temperatures of $\sim 3 \times 10^6$ K, suggesting that photon-starved shocks in which the radiation and gas are not in thermal equilibrium are necessary to explain the temperatures of QPE flares (as argued by \citealt{Linial2023}).

\section{Implications for QPE Models}
\label{sec:dE}

\begin{figure}
    \centering
    \includegraphics[width=\columnwidth]{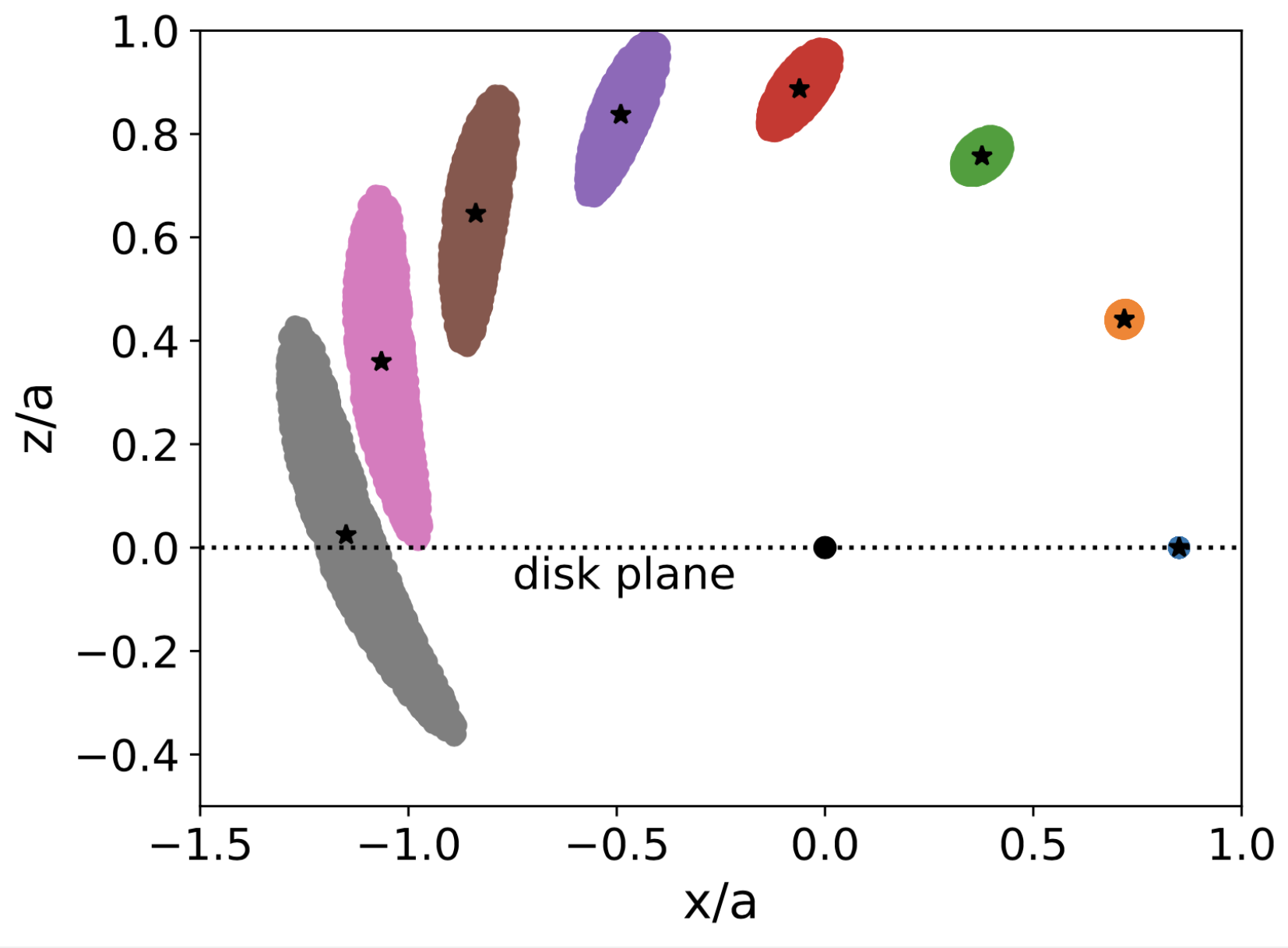}\\   
    \caption{Evolution of the stellar debris produced in a star disk collision, based on numerically integrating an ensemble of test particle orbits.   The gravity of the star is neglected (i.e., we assume that the debris rapidly exits the Hill sphere and then orbits in the potential of the black hole).  Debris clouds are equally spaced in time, distances are in units of the star's semi-major axis $a$, the black hole is located at the origin, and the star's position is marked with a $\star$. Note that $\sim 1/2$ of the debris returns to the disk plane before the star and will collide and shock with the disk.  Calculations are dimensionless but scaled to a $1 M_\odot$ star assuming that the stellar debris is initially moving at $1.5 v_{\rm esc, \star}$ radially away from the star whose orbital velocity is $0.05$c and eccentricity is 0.15.}
    \label{fig:debris_disk}
\end{figure}

The calculations in \S \ref{sec:energy} show that after each star-disk collision, the unbound stellar debris has a spread in velocity $\sim v_{\rm esc,\star}$.  This produces an expanding dense gas cloud around the star with an area significantly larger than that of the star (Fig. \ref{fig:athsims}).   As a result, the cumulative effect of multiple star-disk collisions can be dominated by stellar debris-disk collisions, rather than the direct star-disk collisions.    However, the post-collision dynamics of the unbound stellar debris is not necessarily well-captured by our simulations, which neglect the gravity of the BH. Neglecting the BH's gravity is only a good approximation inside the star's Hill sphere. Gas moving at $v \sim v_{\rm esc,\star}$ will, however, exit the Hill sphere in a small fraction of the orbital period of the star around the black hole, roughly at time $t/P \simeq (2 \pi)^{-1}(v_{\rm esc,\star}/v) (\rt/a)^{1/2}$.   To understand the dynamics of this gas and its interaction with the accretion disk $\sim 1/2$ an orbit later, when the star and unbound stellar debris will again collide with the disk, we turn to analytic estimates motivated by the physical conditions found in the simulations.   We consider two limits in what follows:  gas that escapes the star's Hill sphere as is stretched by the tidal gravity of the BH (Fig. \ref{fig:debris_disk}) and gas that remains within the star's Hill sphere.   In general, the stellar debris can in principle collide multiple times with the disk before it circularizes and is incorporated into the disk.  In what follows we only consider the orbital properties of the stellar debris determined after it is immediately stripped from the star; we motivate this in \S \ref{sec:oneormore} but a full understanding of QPE timing may require following the full circularization process in at least some cases.

In this section, we assume that the star's eccentricity is $e \lesssim 1/2$ and do not distinguish between the star's pericenter distance $r_p$ and the semi-major axis $a$.   

\subsection{Stellar Debris Outside the Hill Sphere}
\label{sec:stretched}

Without loss of generality, we consider a coordinate system such that the star-disk collision happens at $z = y = 0$ but $x \ne 0$ (this is equivalent to $\Omega = \omega = 0$ in terms of Keplerian orbital elements).   The star thus has x-angular momentum $j_x = 0$.   We assume that after the star-disk collision, the unbound stellar debris is distributed over a region $\sim R_\star$ in size surrounding the star and is moving radially away from the star with a speed $v \sim v_{\rm esc,\star}$.  
The gas moves roughly on ballistic trajectories in the frame of the star, until it exits the star's Hill sphere.   The spread in velocity (and hence energy) of the gas relative to the star corresponds to a spread in the semi-major axis once the gas exits the Hill sphere of $\Delta a/a \sim 2 v_{\rm esc,\star}/v_\star$. This can be expressed as
\begin{equation}
\frac{\Delta a}{a} \sim 2^{3/2} \pfrac{a}{\rt}^{1/2} \pfrac{M_\star}{\MBH}^{1/3}
\label{eq:da}
\end{equation}
or
\begin{equation}
\frac{\Delta a}{R_\star} \sim 2^{3/2} \pfrac{a}{\rt}^{3/2} 
\label{eq:da2}
\end{equation}
This spread in the semi-major axis also corresponds to a spread in arrival times of the stellar debris when it returns and collides with the disk of
\begin{equation}
\frac{\Delta t}{P} \sim 3 \sqrt{2} \pfrac{a}{\rt}^{1/2} \pfrac{M_\star}{\MBH}^{1/3}
\label{eq:dt}
\end{equation}

The spread in semi-major axis estimated in equation \ref{eq:da} determines one of the dimensions of the stellar debris stream as it orbits the black hole and later collides with the disk.   The dimension in the other direction is determined by the dispersion in the gas's x-component of angular momentum relative to the BH.   So long as the time for the gas to exit the Hill sphere is small compared to the orbital period, one can show that the x-component of the gas's angular momentum relative to the black hole is roughly conserved even though the gravity is dominated by the star.   The dispersion in the gas's x-angular momentum is thus roughly given by its initial value after the hydrodynamic phase of the interaction is complete, i.e., $\Delta j_x \sim R_\star v_\star$.   Once the gas exits the Hill sphere the value of $j_x$ is of course rigorously conserved.  When the gas returns to $z \sim 0$ and collides with the disk roughly half an orbit later the dispersion in the position of its collision location due to the dispersion in $j_x$ is given by $\Delta y \sim R_\star$.   Combining this estimate with equation \ref{eq:da} we conclude that the total area of the disk intercepted by the stellar debris when it next collides with the disk is 
\begin{equation}
\frac{A_{\rm debris}}{\pi R_\star^2} \sim \pfrac{a}{\rt}^{3/2} \gg 1
\label{eq:debrisarea}
\end{equation}
We have confirmed the estimate in equation \ref{eq:debrisarea} using calculations of test particles orbiting under the influence of stellar and black hole gravity with an initial dispersion in positions and velocities motivated by our hydrodynamic simulations in the previous section; an example of the debris trajectories is shown in Figure \ref{fig:debris_disk}. We note, however, that these calculations are an idealized model for the evolution of the stripped stellar debris in the gravity of the black hole, meant to illustrate the rough geometry the debris obtains. More realistic calculations including the star's gravity and realistic initial conditions of the ram-pressure stripped debris will be important for better quantifying the properties of the stellar debris as it returns and collides with the disk.

There are two key quantities that determine how the stellar debris created in one star-disk collision interacts with the disk in its next collision: (1) the relative surface densities of the stellar debris $\Sigma_{\rm debris} = \Delta M_\star/A_{\rm debris}$ and the disk with which it collides $\Sigma_{\rm disk}$. This determines how quickly the stellar debris decelerates and which component (the lighter one) is shocked to the highest temperatures.   (2) The timescale for the debris stream to mix with the disk via Kelvin-Helmholz and other instabilities (the ``cloud crushing time"; \citealt{Klein1994})
relative to the time for the debris to cross the disk.  In what follows we focus on estimating $\Sigma_{\rm debris}/\Sigma_{\rm disk}$ and leave analysis of the mixing of stellar debris and disk material to future work.  

The surface density of the stellar debris for gas outside the star's Hill sphere can be determined by combining equations \ref{eq:fit} and \ref{eq:debrisarea}:
\begin{equation}
\frac{\Sigma_{\rm debris}}{\Sigma_{\rm disk}} \sim 0.03 \pfrac{\rt}{a}^{5/2} \pfrac{M_{\rm BH}}{M_\star}^{2/3} \left[1 + \frac{H_{\rm disk}}{R_\star}\right]^{-1}
\label{eq:Sigratio}
\end{equation}
Equation \ref{eq:Sigratio} shows that are in principle two regimes possible for star-disk transients depending on $\rt/a$ and $\MBH/M_\star$:   $\Sigma_{\rm debris} \lesssim \Sigma_{\rm disk}$ or $\Sigma_{\rm debris} \gtrsim \Sigma_{\rm disk}$. When $\Sigma_{\rm debris} \ll \Sigma_{\rm disk}$, the stellar debris is spread out over such a large area that it does not significantly perturb the structure of the disk.  In this regime, the collision between the stellar debris and the disk shocks the stellar debris to a high speed $\sim v_\star$ potentially producing a flare of radiation.   In addition, since the disk structure is not strongly perturbed by the stellar debris, the star-disk collision is a second possible source of a flare of radiation.  In general, the shocked stellar debris mass and energy will exceed the disk mass directly shocked by the star. All things being equal, the stellar debris, not the star-disk collision, is thus likely to dominate the emission.  However, it is not clear that each shock component will emit with similar radiative efficiency (or at similar photon energies) so more detailed calculations of the radiation are required to assess which component dominates.

In the alternative regime $\Sigma_{\rm debris} \gg \Sigma_{\rm disk}$, the star-disk collision is important primarily for unbinding mass from the star; the dominant source of radiation is likely to be the shock between the stellar debris and the disk, which accelerates the disk material to high temperatures and speeds of $\sim v_\star$.  The stellar debris is more important than the star itself because the debris occupies a larger area than the star and has shocked the disk prior to the star's arrival at the next collision point.  We show below in \S \ref{sec:QPE} that observed QPEs are largely in this regime of $\Sigma_{\rm debris} \gtrsim \Sigma_{\rm disk}$.

\subsection{Gas Within the Star's Hill Sphere}
\label{sec:hillsphere}

The energy distribution of the stellar debris in Figure \ref{fig:energy_dist} shows that there is a significant amount of mass that is marginally bound to the star after each collision.   It is plausible that some (or most) of this gas does not escape the star's Hill sphere and instead remains within it.  The size of the Hill sphere is $r_H \simeq R_\star (a/\rt)$ so that in this regime the area of the debris when it returns and collides with the disk is given by
\begin{equation}
\frac{A_{\rm debris}}{\pi R_\star^2} \sim \pfrac{a}{\rt}^{2}.
\label{eq:debrisarea2}
\end{equation}
The relative surface density of the stellar debris and disk is then given by
\begin{equation}
\frac{\Sigma_{\rm debris}}{\Sigma_{\rm disk}} \sim 0.03 \pfrac{\rt}{a}^{3} \pfrac{M_{\rm BH}}{M_\star}^{2/3} \left[1 + \frac{H_{\rm disk}}{R_\star}\right]^{-1}
\label{eq:Sigratio2}
\end{equation}
while the relative volume density of the debris and disk gas is given by
\begin{equation}
\begin{aligned}
\frac{\rho_{\rm debris}}{\rho_{\rm disk}} \sim & \, \, 0.03 \pfrac{\rt}{a}^{4} \pfrac{M_{\rm BH}}{M_\star}^{2/3} \pfrac{H_{\rm disk}}{R_\star} \\ 
& \times \left[1 + \frac{H_{\rm disk}}{R_\star}\right]^{-1}.
\label{eq:rhoratio2}
\end{aligned}
\end{equation}
The dispersion in arrival time of the stellar debris, as it returns and impacts the disk, is set by the size of the Hill sphere, with
\begin{equation}
\frac{\Delta t}{P} \sim \frac{r_H}{a} \sim \pfrac{M_\star}{\MBH}^{1/3}
\label{eq:dt2}
\end{equation}

\subsection{Implications for QPEs}
\label{sec:QPE}

We now discuss the implications of our results for models of QPEs produced by star-disk collisions; none of what follows applies if the impactor is a compact object (as in, e.g., \citealt{Franchini2023}). 
The most qualitatively important feature of our results is that the
stellar debris subtends a larger area than the star when it returns to collide with the disk (eqs \ref{eq:debrisarea} \& \ref{eq:debrisarea2}). In many cases, `star-disk' QPE models are thus better viewed as stellar debris-disk circularization shocks than star-disk collisions.  Two other consequences follow from this:   (2) The dispersion in energy in the stellar debris inevitably leads to a dispersion in collision times between the stellar debris and the disk (eqs \ref{eq:dt} and \ref{eq:dt2}).  This may be important for setting the duration of QPE flares.  (3) All else being equal, debris-disk collisions are more likely to produce a one-sided plume of shocked disk material rather than a symmetric ejection of disk material. This is because the symmetric ejection of disk material is realized only when the size of the impactor is small compared to the thickness of the disk; the larger size of the stellar debris cloud makes this more difficult to realize:\footnote{We thank Itai Linial and Brian Metzger for emphasizing this point to us.}  the two panels in Figure \ref{fig:disk_symmetry} demonstrate this point explicitly.  

For concreteness, we consider a main-sequence star with a mass of $1 M_{\star,1} M_\odot$ orbiting a BH of mass $M_{\rm BH} = 10^6 M_{\rm BH,6} M_\odot$ with an orbital period of $P = 10 P_{10}$ hrs (if there are two QPE flares per orbit, as in models to reproduce the alternating long-short time between flares in GSN 069 and eRO-QPE2 \citep{Miniutti2019,Miniutti2023,Arcodia2021}, then the orbital period $P$ is roughly twice the QPE recurrence time $t_{\rm QPE}$).   The ratio of the semi-major axis of the star's orbit to the tidal radius is then given by
\begin{equation}
\frac{a}{\rt} \simeq 2.3 \, M_{\star,1}^{-1/2} \, P_{10}^{2/3}
\label{eq:aoverrt}
\end{equation} 

The shortest period QPEs have recurrence times of $\sim$2.4 hrs \citep{Arcodia2024}.   Accounting for the possibility of two flares per orbit and main sequence stars with masses as low as 0.1 $M_{\odot}$, this implies that known QPEs are consistent with $a/r_t \gtrsim 4.5$.   At this separation, our neglect of tidal heating of the star is plausible based on the calculations in \citet{Linial2024c}. The neglect of tidal heating of the star is better for longer recurrence time systems. In addition, the calculations in \S \ref{sec:stellarmassloss} show that matter at distances of 4.5 $R_\star$ is already unbound from the star in our simulations of star-disk collisions that do not include the gravity of the BH. Most QPEs are thus plausibly in the regime where the BH's tidal forces are important for the orbital evolution of the debris (as in Figure \ref{fig:debris_disk}), but not necessarily for the initial stripping of stellar material. This argument assumes that the shortest recurrence time QPEs are from very low mass stars. If they are instead from ~1 $M_{\odot}$ stars with orbital periods of P $\sim$ 5 hrs, then $a/r_t \sim 1.4$ and the star could not have survived that close to the black hole for any prolonged period of time; this is consistent with the need for low mass stars at least in the shortest recurrence time QPE systems.

\subsubsection{Lifetime of the QPE-Emitting Phase}
\label{sec:lifetime}
We now recast equation \ref{eq:fit} in a form that is more easily related to QPE observables.  To do so, we adopt the radiation model of \citep{Linial2023,Franchini2023,Tagawa2023} in which the shocked disk material produces the QPE flares (what follows is not sensitive to details of the radiation model, just this basic physical picture).  We assume that the ejecta mass in each QPE flare $M_{\rm ej} = f_M M_{\rm disk}$ where $M_{\rm disk}$ is (as in \S \ref{sec:massloss}) the amount of disk mass that would have been shocked by the star alone and $f_M > 1$ is possible because of the larger area of the stellar debris; the simplest estimate is $f_M \sim A_{\rm debris}/\pi R_\star^2$ as calculated in Equations \ref{eq:debrisarea} and \ref{eq:debrisarea2}, but in principle $f_M$ could be yet larger if there is efficient mixing of the stellar debris and shocked disk during their collision.  The energy of shocked disk material produced in each collision is then $E_{\rm ej} = 1/2 M_{\rm ej} v_\star^2$ and we assume that the observed radiated energy in each QPE flare is $E_{\rm flare} = \eta E_{\rm ej}$ where $\eta \lesssim 1$ is the radiative efficiency of the flare; $\eta \sim 0.1$ in \citet{Linial2023}.   Equation \ref{eq:fit} can then be expressed as
\begin{equation}
\frac{\Delta M_\star}{M_\star} \simeq \frac{0.06}{f_M \eta} \frac{E_{\rm flare}}{GM_\star^2/R_\star} \left[1 + \frac{H_{\rm disk}}{R_\star}\right]^{-1}
\label{eq:dMfitobs}
\end{equation}
The lifetime of the star in the QPE-emitting phase set by mass stripping due to the repeated star disk interactions  is thus $t_{\rm lifetime} \simeq t_{\rm QPE} M_\star/\Delta M_\star$, i.e., 
\begin{equation}
\begin{aligned}
t_{\rm lifetime} & \sim 10 \,{\rm yr} \, f_M \eta \pfrac{t_{\rm QPE}}{10 \, {\rm hr}} \pfrac{E_{\rm flare}}{10^{45.7} \, {\rm erg}}^{-1} M_{\star,1}^{1.2} \\ & 
\hspace{2cm} \times \left[1 + \frac{H_{\rm disk}}{R_\star}\right] 
\label{eq:lifetime} \\ & 
\sim 200 \, {\rm yr} \, \eta 
\pfrac{t_{\rm QPE}}{10 \, {\rm hr}}^{7/3} \pfrac{E_{\rm flare}}{10^{45.7} \, {\rm erg}}^{-1} M_{\star,1}^{0.2} \\ & 
\hspace{2cm}  \times \left[1 + \frac{H_{\rm disk}}{R_\star}\right]
\end{aligned}
\end{equation}
where we have assumed $R_\star \propto M_\star^{0.8}$ for main-sequence stars and have scaled the flare radiated energy and recurrence time to values appropriate for GSN 069 \citep{Miniutti2019}.   In the second expression in equation \ref{eq:lifetime} we have assumed $P = 2 t_{\rm QPE}$ (2 flares per orbit) and have
used $f_M \sim A_{\rm debris}/\pi R_\star^2$ from equation \ref{eq:debrisarea2}, appropriate for gas filling the Hill sphere (though the total mass contributing to the QPE flare may be larger than this if there is efficient debris-disk mixing).  The results for gas that is tidally stretched (eq. \ref{eq:Sigratio2}) are similar to within a factor of a few.   

Equation \ref{eq:lifetime} is best interpreted as an upper limit on the lifetime of the QPE-emitting phase.   The true lifetime of the star due to star-disk collisions could in principle be either shorter or longer than this depending on how the accretion disk with which the star interacts evolves on a timescale of $\sim t_{\rm lifetime}$.   
Equation \ref{eq:lifetime} shows that the lifetime of the QPE-emitting phase in models of GSN 069 is remarkably short unless $f_M \eta \gtrsim 1$, i.e. unless the radiative efficiency is relatively high and the disk is shocked primarily by the stellar debris. 
In general, $f_M \gtrsim 1$ will also lead to higher radiative efficiency, further contributing to a longer lifetime predicted by equation \ref{eq:lifetime}:  the reason is that the larger collision area will mean less adiabatic losses as the debris cloud expands to the point where photon diffusion is rapid and the flare emission escapes.   Larger $H_{\rm disk}/R_\star$ also increases the lifetime because the stellar mass-loss per collision relative to the local disk mass is smaller in this regime (eq. \ref{eq:fit}). A larger value of $H_{\rm disk}/R_\star$ is also favored to explain the long-short timing pattern in GSN 069 \citep{Linial2023}.

The observed system to date for which our predicted lifetime is the shortest is eRO-QPE2 \citep{Arcodia2021}, which has $t_{\rm QPE} \sim 2.5$ hr and $E_{\rm flare} \sim 10^{45}$ erg, implying $t_{\rm lifetime} \sim 40 \eta \, (1+ H_{\rm disk}/R_\star)$ yrs if equation \ref{eq:debrisarea2} for $f_M$ is appropriate.  
One interesting tension is that the quiescent non-flaring luminosity of eRO-QPE2 is low ($L \sim 3 \times 10^{41} \ergss$), much lower than would be expected if the star is being destroyed on a timescale of decades.   This could imply that the system is not in a steady state, and that the stellar mass ablated per orbit is piling up in the disk, rather than accreting efficiently onto the central BH.   If correct, this suggests that eRO-QPE2 may undergo an outburst of luminous accretion in the future.

\subsubsection{Implications for QPE Timing}

We now consider the implications of our estimates of the stellar debris area and density when it impacts the disk.   We focus on the case in which the stellar debris remains within the Hill sphere, primarily for convenience given the simple expressions for the debris density and geometry in this case.
Equation \ref{eq:Sigratio2} and \ref{eq:rhoratio2}, which determine how efficiently the stellar debris and disk mix and decelerate during their collision, can be written as 
\begin{equation}
\frac{\Sigma_{\rm debris}}{\Sigma_{\rm disk}} \sim 30 \, M_{\star,1}^{5/6} P_{10}^{-2} M_{\rm BH,6}^{2/3} \left[1 + \frac{H_{\rm disk}}{R_\star}\right]^{-1}
\label{eq:Sigobs}
\end{equation}
and
\begin{equation}
\frac{\rho_{\rm debris}}{\rho_{\rm disk}} \sim 10 \, M_{\star,1}^{4/3} P_{10}^{-8/3} M_{\rm BH,6}^{2/3} \left[\frac{H_{\rm disk}/R_\star}{1 +  H_{\rm disk}/R_\star}\right]
\label{eq:rhoobs}
\end{equation}

The recurrence time of observed QPEs range from $t_{\rm QPE} \simeq 2.4-20$ hrs \citep{Miniutti2019,Arcodia2021,Arcodia2024}.  Even accounting for the possibility of $P \simeq 2 t_{\rm QPE}$, i.e., 2 flares per orbit, equation \ref{eq:Sigobs} implies that most observed QPEs to date are in the regime where $\Sigma_{\rm debris} \gtrsim \Sigma_{\rm gas}$ and $\rho_{\rm debris} \gtrsim \rho_{\rm disk}$, although only marginally so for eRO-QPE1 and eRO-QPE3, which have recurrence times of $t_{\rm QPE} \simeq 20$ hr (we return to this case below).  

In the regime of $\Sigma_{\rm debris} \gtrsim \Sigma_{\rm gas}$ and $\rho_{\rm gas} \gtrsim \rho_{\rm disk}$ appropriate for shorter period QPEs, the dominant interaction likely to power observed emission is between the stellar debris created in a previous collision and the accretion disk, not the direct collision of the star with an otherwise unperturbed disk.   As in the star-disk collision case \citep{Linial2023}, the most likely source of QPE emission, in this case, is the shocked disk material because it is of lower density and heated to higher temperatures than the shocked stellar debris.  Our results demonstrate, however, that the shocked disk mass is likely significantly larger than assumed in previous work because of the larger cross-section of the stellar debris relative to the bound star.

One of the most striking features of observed QPEs is the varied timing behavior.   Several systems, e.g., GSN 069 and eRO-QPE2, show a pattern in which the time between consecutive flares differs by $\sim 10 \%$ while in other systems, e.g. eRO-QPE1, the time between consecutive flares is much more erratic (in eRO-QPE2 this long-short pattern of recurrence times was present initially but not in subsequent observations several years later; \citealt{Arcodia2024b}).  The alternating long-short recurrence time has been elegantly explained in models with two flares per orbit due to star-disk collisions, with the difference in the time between flares related to the eccentricity of the orbit \citep{Xian2021,Linial2023,Franchini2023}.   

Our results provide a possible explanation for the diversity in QPE timing behavior.  It is natural to expect that the more complex hydrodynamics of stellar debris-disk collisions (relative to star-disk collisions) can produce a range of outcomes and timing properties.  
We suggest that the long-short pattern of recurrence times in GSN 069 and eRO-QPE2 is likely to be present either if $\Sigma_{\rm debris} \gg \Sigma_{\rm disk}$ and perhaps if $\Sigma_{\rm debris} \ll \Sigma_{\rm disk}$, but not if $\Sigma_{\rm debris} \sim \Sigma_{\rm disk}$.

If $\Sigma_{\rm debris} \gg \Sigma_{\rm disk}$ the stellar debris dominates over the star in determining the shocked disk properties and thus the QPE observables.  Because of the large density contrast and short interaction time, the debris-disk collision does not produce significant mixing between the shocked disk and stellar debris but instead simply ejects the shocked disk at a speed $\sim v_\star$.   Nonetheless, the timing of the stellar debris' impact with the disk is still set by the star's orbit, potentially producing the long-short pattern of recurrence times observed.   This is likely to be the case so long as each stellar debris stream only produces a single QPE flare (otherwise there could be many flares due to the different debris-disk interactions all with different properties); we discuss this in \S \ref{sec:oneormore}.   

In the alternative regime of $\Sigma_{\rm debris} \ll \Sigma_{\rm disk}$, the stellar debris does not significantly impact the disk.  Instead,  observed flaring is likely to be produced by the stellar debris being shocked to high speed and temperature by its collision with the disk.  Star-disk collisions will also eject disk material at high temperature and speed, but with less overall mass and energy than the stellar debris.  Detailed radiation calculations are needed to understand which shocked component dominates in different parts of the spectrum. We have shown that the regime $\Sigma_{\rm debris} \ll \Sigma_{\rm disk}$ is not, however, realized in the current sample of QPEs, though it would potentially apply to even longer period QPEs (eq. \ref{eq:Sigobs}).

We speculate that the more complex timing behavior seen in, e.g., eRO-QPE1, may be a consequence of $\Sigma_{\rm debris} \sim \Sigma_{\rm disk}$.   The scalings in equation \ref{eq:Sigobs} imply that longer orbital periods and lower mass stars are more likely to have $\Sigma_{\rm debris} \sim \Sigma_{\rm disk}$, consistent with eRO-QPE1's longer recurrence time making it distinct from other QPE sources (eRO-QPE3 has a similar recurrence time but has been studied less so far; \citealt{Arcodia2024}).  In the regime $\Sigma_{\rm debris} \sim \Sigma_{\rm disk}$, the debris disk collisions will partially mix leading to dispersion in how much mass and energy is accelerated to high speed in each collision depending on the exact hydrodynamics of the debris-disk mixing.  In addition, both the shocked disk material and stellar debris are likely to be shocked to high speeds and temperatures; if mixing is only partial, these two components may contribute separately to the observed emission.  These effects - unique to the $\Sigma_{\rm debris} \sim \Sigma_{\rm disk}$ regime - will all produce more flare-to-flare variability.   At longer orbital periods, the star's Hill sphere is also larger making it perhaps more plausible that stellar debris is both unbound from the Hill sphere and retained within it.   These two distinct components could each produce QPEs, further contributing to the variability in flare properties and timing.  If our suggestion regarding the timing properties of eRO-QPE1 is correct, further observations of eRO-QPE3 are likely\footnote{This assumes that the stellar mass and BH mass in eRO-QPE3 are similar to those in eRO-QPE1, which is plausible but non-trivial to accurately test.} to show diverse timing similar to that of eRO-QPE1 while even longer recurrence time QPEs may show more regular timing because they are in the regime $\Sigma_{\rm debris} \ll \Sigma_{\rm disk}$.

\subsubsection{Multiple QPE Flares Per Stellar Debris Stream?}
\label{sec:oneormore}

The estimates in \S \ref{sec:stretched} and \ref{sec:hillsphere} only apply the `first' time a recently stripped stellar debris cloud returns and impacts the disk.  During this collision, however, the stellar debris will be further heated and will thus expand to collide with an even larger area of the disk.  
Following the full circularization process is beyond the scope of this paper, but a simple estimate shows that it is likely that the debris rapidly heats up even if initially $\Sigma_{\rm debris} \gtrsim \Sigma_{\rm disk}$.   Assuming that $\rho_{\rm debris} > \rho_{\rm disk}$, the speed of the shock into the stellar debris during the debris-disk collision is $\Delta v \simeq v_\star (\rho_{\rm disk}/\rho_{\rm debris})^{1/2}$.    Using equations \ref{eq:rhoratio2} and \ref{eq:aoverrt} we estimate that
\begin{equation}
\frac{\Delta v}{v_{\rm esc,\star}} \sim 10 \, M_{\star,1}^{-3/4} P_{10} \left[\frac{H_{\rm disk}/R_\star}{1 + 0.5 H_{\rm disk}/R_\star}\right]^{-1/2}
\label{eq:dv}
\end{equation}
We have compared $\Delta v$ in equation \ref{eq:dv} to $v_{\rm esc,\star}$ because the latter sets both whether stellar debris can leave the Hill sphere and the energy spread of the stellar debris produced as it is stripped from the star in the star-disk collision (Fig. \ref{fig:energy_dist}).   Equation \ref{eq:dv} shows that in general the energy spread introduced in the disk-debris collision is significantly larger than that produced in the initial stripping of the stellar debris. Equation \ref{eq:dv} implies that in its next (i.e., second) collision with the disk, the stellar debris is likely to be spread out over such a large area that it is efficiently heated to even higher temperatures and velocities, generating an expanding cloud with a size similar to that of the star's orbit. It is possible that in some circumstances this second collision itself produces another QPE, but more than two per stellar debris cloud appears very unlikely. Alternatively, this `2nd' collision could produce additional low amplitude variability similar to the QPO seen in GSN 069 \citep{Miniutti2023}. More detailed calculations following this hydrodynamics would be very valuable.    

\section{Summary and Discussion} \label{sec:summary}
Stars near SMBHs can undergo significant orbital and structural changes due to tidal forces and collisions with an accretion disk. Understanding these processes may be critical to understanding the emerging class of repeating nuclear transients in galactic nuclei, including quasi-periodic eruptions (QPEs) and partial tidal disruption events (pTDE). With this motivation in mind, we have studied the physics of star-disk collisions in (primarily 2D) \texttt{Athena++} hydrodynamic simulations, focusing on the impact on the star.   We also provide an analytic model of aspects of star-disk collisions inspired by the simulations.   Because our focus is on the hydrodynamics of the star and stellar debris, our results do not apply to models in which repeating nuclear transients are produced by compact object-accretion disk collisions.  Our simulations focus on a 1 $M_\odot$ star that is initially a $\gamma=5/3$ polytrope; we argue in \S \ref{sec:othermass} that our results can be scaled (at the factor of few level) to other low-mass stars, as we have done in equations \ref{eq:1stfit}, \ref{eq:fit}, and throughout \S \ref{sec:dE}.

Our primary results and their implications include:

(i) For stars with orbital periods of order hours to days, such as in QPE models, radiative losses after star-disk collisions are not dynamically important because the thermal time of the shocked gas is much longer than the orbital period ($t_{\rm th} \gg P$, see \S \ref{sec:thermal}).  The entire outer layers of the star are thus shock-heated by repeated collisions with the disk. We have accounted for this in our simulations by carrying out a sufficient number of star-disk collisions so that the stripped mass per collision reaches an approximate steady state.  On the other hand, for longer-period orbits (e.g., pTDE candidates such as ASASN-14ko) the shocked outer layers of the star can at least partially cool and contract between collisions. 

(ii) In the absence of significant cooling, shock heating causes the bound stellar envelope to expand to $\sim 2-3 R_{\star}$ after only a few collisions, with a cloud of unbound stellar debris yet further from the star.  The collision cross-section for the star to interact with the disk thus increases significantly as a result of repeated star-disk collisions (\S \ref{sec:massloss} and Figs \ref{fig:athsims} \& \ref{fig:disk_symmetry}).

(iii) The mass-loss per star-disk collision after multiple collisions is nearly 20 times larger than the mass-loss produced in the `first' collision between a disk and an unperturbed star (\S \ref{sec:stellarmassloss}).  This is because the shock-heated envelope of the star is much easier to unbind than an initially cold stellar envelope. After multiple collisions, the star-disk interactions reach a rough statistical steady state in which each collision unbinds a significant amount of previously shocked envelope while at the same time deeper layers are shocked and expand to replenish the outer parts of the bound stellar envelope (Fig. \ref{fig:masszones}).  A key consequence of the greater mass loss after multiple collisions is that the lifetime of the star in models in which QPEs are produced by star-disk collisions is significantly shorter than previously predicted.   We provide fitting functions for the stellar mass-loss per collision valid in both the multiple-collision (eq. \ref{eq:fit}; no cooling) and first-collision (eq. \ref{eq:1stfit}; unperturbed star) regimes.  The latter is likely more appropriate for stars on long-period orbits $\gtrsim$ yr (eq. \ref{eq:thermal}).  

(iv) The lifetime of the QPE-emitting phase in models in which QPEs are produced by star-disk collisions has an upper limit set by the time between collisions $t_{\rm QPE}$ and the mass stripped per collision, i.e. $t_{\rm lifetime} \lesssim t_{\rm QPE} M_\star/\Delta M_\star$ (\S \ref{sec:lifetime}).  We show analytically how our simulation results can be expressed as a QPE lifetime that depends on QPE observables as well as a few theoretical parameters of star-disk collision models (eq. \ref{eq:lifetime}).  This estimate assumes that the shocked disk dominates the observed QPE flares (as in, e.g., \citealt{Linial2023,Franchini2023,Tagawa2023}) and does not apply if the flares are instead produced by the shocked stellar debris.  Our lifetime estimate is, of course, also uncertain at the factor of a few level given the idealized nature of our simulations (polytropic initial stellar model; overly-idealized disk structure, etc). It is nonetheless striking that we find that QPE lifetimes are at most $\sim 1000$ yrs and could be as short as a few decades depending on the radiative efficiency of QPE flares and the thickness of the disk with which they collide. This range of QPE lifetimes follows from the observed diverse QPE recurrence times, and is consistent with lifetime estimates of $\sim 100-1000$ years from \citet{Lu2023} and \citet{Linial2023}. Given the current uncertainties on the volumetric rates of QPEs \citep{Arcodia2024c}, our estimated lifetime is also consistent with formation mechanisms of stellar extreme-mass-ratio-inspirals on mildly eccentric orbits via Hills breakup of binaries (see detailed discussions in \citealt{Linial2023}). This limit on the lifetime only applies to the QPE phase as currently observed; if the underlying accretion disk with which the star collides evolves on a timescale $\lesssim t_{\rm lifetime}$ (as in TDE models; \citealt{Linial2023}) the lifetime will change accordingly.   Of observed QPEs, we predict the shortest QPE-emitting lifetime for eRO-QPE2 of only a few decades (with, we stress again, significant theoretical error bars). The quiescent (between flare) luminosity of eRO-QPE2 is not compatible with the accretion of a stellar mass on this timescale so the system is likely not in a steady state, with the stripped stellar mass piling up in an accretion disk.  This strongly suggests that eRO-QPE2 will undergo an accretion outburst in the future if its QPEs are indeed produced by star-disk collisions (there is, however, no evidence of significant secular evolution in eRO-QPE2 to date; \citealt{Arcodia2024b}).  

(v) Unbound stellar debris stripped from the star after each star-disk collision has a velocity spread of order the stellar escape speed $v_{\rm esc,\star}$; this produces an expanding debris cloud around the star (Fig. \ref{fig:athsims}). Asymmetry in the orbital angular momentum and energy removed by the debris cloud could contribute to the finite eccentricity of the stars' orbit, even if the orbit was circular prior to the onset of star-disk collisions; this would be valuable to study in future work. Our simulations only include the stellar gravity, not that of a BH, and so only capture the dynamics of gas that remains within the star's Hill sphere. We use our simulation results, together with analytic estimates of the properties of the stellar debris cloud, to quantify the interaction between the debris and disk $\sim 1/2$ an orbit later when the stellar debris collides and shocks with the disk (\S \ref{sec:stretched}).   The properties of this interaction are governed to a large extent by the ratio of the stellar debris surface density $\Sigma_{\rm debris}$  to the surface density of the disk with which it collides $\Sigma_{\rm disk}$. We show analytically that this ratio depends most sensitively on the orbital period of the orbiting star, with weaker dependence on the stellar and BH mass and disk thickness (eq. \ref{eq:Sigobs}); this suggests that orbital period (and hence QPE recurrence time) may be the most important parameter determining the overall properties of a QPE source.

Currently observed shorter period QPEs, such as eRO-QPE2 and GSN 069, are in the regime $\Sigma_{\rm debris} \gg \Sigma_{\rm disk}$.   In this regime the stellar debris occupies an area larger than that of the star and thus shocks a larger disk mass than the star alone would have done (the stellar debris also shocks the disk prior to the star doing so; Fig. \ref{fig:debris_disk}).  When 
$\Sigma_{\rm debris} \gg \Sigma_{\rm disk}$, QPEs are thus not produced by star-disk collisions, but can instead be produced by the disk material shocked and heated to high temperatures in stellar debris-disk collisions.  Nonetheless, the timing of the debris-disk collisions is still set by the stellar orbit, consistent with previous explanations for the alternating long-short pattern of recurrence times in eRO-QPE2 and GSN 069.

No currently observed QPEs are in the regime of $\Sigma_{\rm debris} \ll \Sigma_{\rm disk}$ but longer recurrence time systems, if they exist, would be (eq. \ref{eq:Sigobs}).  In the regime $\Sigma_{\rm debris} \ll \Sigma_{\rm disk}$ the stellar debris has little impact on the disk with which it collides.  Instead, it is the stellar debris that is shocked to high speeds and temperatures by the collision with the disk, potentially producing a QPE in much the same way that the shocked disk material does for $\Sigma_{\rm debris} \gg \Sigma_{\rm disk}$.  In this regime, the star also collides with a disk that is relatively unperturbed by its interaction with the stellar debris, as in previous models of QPEs from star-disk collisions \citep{Linial2023}.  However, the energy in the disk material shocked by the star is significantly smaller than the energy in the stellar debris shocked by the debris-disk interaction.   This suggests that the star-disk collisions will be subdominant in producing observed flares, but calculations of the radiation produced by each component are required to be certain of this conclusion. If the stellar debris-disk collision indeed produces a QPE when $\Sigma_{\rm debris} \ll \Sigma_{\rm disk}$, the shocked stellar debris will be one-sided and thus it is likely that only one collision per orbit will be observable (unless we observe the system nearly edge-on).

We speculate that in the regime $\Sigma_{\rm debris} \sim \Sigma_{\rm disk}$, which includes systems with $t_{\rm QPE} \sim$ day such as eRO-QPE1 and eRO-QPE3, that the QPE lightcurves, duration, and recurrence times, are much more influenced by the hydrodynamics of the debris-disk collision than in other systems.  For example, both the shocked stellar debris and shocked disk material may contribute to observed QPEs depending on the exact amount of mixing during the debris-disk collision.    This more complex hydrodynamics may account for the larger timing variations in eRO-QPE1 relative to other systems \citep{Arcodia2024}.   A larger sample of QPEs can test our prediction that only intermediate period systems with periods of $\sim$ day will show such irregular timing (with both shorter and longer period systems showing more regular timing).

In this paper, we have not attempted to directly model the radiation produced in stellar debris-disk collisions. It is possible that only certain regimes of $\Sigma_{\rm debris}/\Sigma_{\rm disk}$ have the right physical conditions for the radiation produced during the collision to outshine the underlying disk in the soft X-rays, producing a QPE.  We do find that the temperature of both the shocked stellar and disk material (Figs. \ref{fig:athsims} \& \ref{fig:disk_symmetry}) in our simulations is too low to explain QPE flares.  This is consistent with \citet{Linial2023}'s argument that photon-starved shocks in which the gas and radiation are out of thermal equilibrium are necessary to explain the temperatures of QPE flares.   

(vi) In our model in which stellar debris-disk interactions produce QPEs, the dispersion in arrival time of the stellar debris to the disk (eqs \ref{eq:dt} and \ref{eq:dt2}) may be important for setting the duration of QPE flares.   The linear relation between arrival time dispersion and orbital period for gas within the star's Hill sphere (eq. \ref{eq:dt}) is tantalizingly similar to the observed correlation between flare duration and recurrence time in QPEs \citep{Arcodia2024}.  
The larger area of the stellar debris (relative to the star itself) also impacts the geometry of the shocked disk material.  As Figure \ref{fig:disk_symmetry} shows, stellar debris is more likely to produce a one-sided plume of shocked disk material than a star-disk collision, which can produce two-sided plumes of disk ejecta when the disk is significantly thicker than the size of the star.  Two-sided plumes of disk ejecta have been invoked to explain the alternating long-short pattern of recurrence times in GSN 069 and eRO-QPE2 \citep{Xian2021,Linial2023, Franchini2023}.   We stress, however, that in our simulations the Hill sphere is effectively infinitely large (no BH gravity) so we may not correctly capture the size of the stellar debris cloud. For realistic parameters, the linear size of the debris cloud in GSN 069 and eRO-QPE2 is probably within a factor of $\sim 3$ of the size of the star (set by eq. \ref{eq:aoverrt}), which is within the uncertainty in theoretical models of the thickness of accretion disks in these sources; stellar debris-disk collision models thus still appear viable for producing an alternating long-short pattern of recurrence times via two-side plumes of disk ejecta produced in stellar debris-disk collisions.

(vii) In principle stellar debris-disk interaction could produce multiple large-amplitude flares during the course of debris circularization.  We argue analytically, however, that the energy spread in the stellar debris increases dramatically during its first collision with the disk (\S \ref{sec:oneormore} and eq. \ref{eq:dv}).   This strongly suggests that stellar debris produced in star-disk collisions will produce at most two, and plausibly just one, large amplitude flare (a QPE) before being spread out over such a large area that it cannot produce a short duty cycle change in emission.   More detailed analytic calculations and hydrodynamic simulations that follow the full circularization process would be very valuable for elucidating this aspect of star disk collisions.  

A limitation of our study is that we do not include the gravity of the BH in our simulations, but only that of the star.  This implies that the stellar debris cloud we find (e.g., Fig \ref{fig:athsims}) is not quantitatively accurate once the stellar debris has expanded outside the star's Hill sphere.   We used analytic arguments to estimate the stellar debris properties for unbound gas that is tidally stretched by the BH's gravity (Fig. \ref{fig:debris_disk} and \S \ref{sec:stretched}). We also showed numerically that the mass of unbound stellar debris produced in later star disk collisions is not sensitive to exactly how much of the previously unbound stellar debris the disk interacts with (Fig. \ref{fig:convergence}).  More detailed calculations self-consistently including the star and BH gravity would nonetheless be very valuable.   These calculations would also reveal whether multiple debris-disk shocks due to gas both within and outside of the star's Hill sphere can contribute to the complex flare timing in systems such as eRO-QPE1.

\section*{Acknowledgments}
We thank Riccardo Arcodia, Xiaoshan Huang, Itai Linial, Brian Metzger, Selma de Mink, and Norm Murray for useful conversations, and Riccardo, Itai, and Brian for helpful comments on an initial draft.   We are grateful to Lars Bildsten, Logan Pruest, and Gabe Kumar for sharing their \texttt{Athena++} equation of state for gas and radiation pressure.   This work was supported in part by a Simons Investigator grant from the Simons Foundation (EQ).   This research benefited from interactions at workshops funded by the Gordon and Betty Moore Foundation through grant GBMF5076 and through interactions at the Kavli Institute for Theoretical Physics, supported by NSF PHY-2309135.

\begin{appendix}

\section{Stellar Mass Loss in the First Collision}

\label{sec:firstcoll}

In the star-disk interactions of interest in this paper, the duration of the star's passage through the disk $t_{\rm cross} \simeq 2 H_{\rm disk}/v_\star$ can be significantly shorter than the star's dynamical time, suppressing the depth the shock reaches and thus the mass stripped from the star.   This effect is particularly prominent for small disk scale heights in the first star-disk collision where the star's structure is that of an unperturbed $n = 3/2$ polytrope (Fig. \ref{fig:1st_hit}).   Here we estimate the depth the shock reaches as a function of $t_{\rm cross}/t_{\rm sh}$ to qualitatively explain the trends in Figure \ref{fig:1st_hit} at small disk scale-height.

The time required for the shock to reach the depth where $p \sim p_{\rm ram}$ is given by equation \ref{eq:tsh} for an $n = 3/2$ polytrope.  This follows from the fact that the pressure as a function of depth $\delta r$ is $p \propto \delta r^{5/2}$ for an n = 3/2 polytrope and the sound speed $c_s \propto \delta r^{1/2}$.
To estimate the mass shocked when $t_{\rm cross} \lesssim t_{\rm sh}$ we start with the fact that at time t the shock will have reached a depth $\delta r$ such that $\delta r \sim v_s(\delta r) t$ where $v_s \sim \sqrt{p_{\rm ram}/\rho(\delta r)}$ is the speed of the shock at the same depth.   For an $n = 3/2$ polytrope the density near the surface scales as $\rho \propto (\delta r)^{3/2}$.    This leads to the depth of the shock as a function of time $t$ being given by 
$\delta r/R_* \sim (p_{\rm ram}/{p_\star})^{2/7 } (t/t_{dyn})^{4/7}$  The energy supplied to the star when it has reached a depth $\delta r$ is $\sim 4 \pi R_\star^2 p_{\rm ram} \delta r$ so that the energy supplied to the stellar envelope for $t_{\rm cross} \lesssim t_{\rm sh}$ is smaller by $\sim (t_{\rm cross}/t_{\rm sh})^{4/7}$ than its asymptotic value when the shock is on for long times.   Even after the external ram pressure subsides at $t \sim t_{\rm cross}$, the shock will continue to propagate into the stellar interior, stalling when the shock has decelerated to the local sound speed, which corresponds to a depth satisfying $E(t_{\rm cross}) \sim M_{\rm ex} c_s^2$, where $M_{\rm ex}$ is the mass exterior to a given location in the star.   Using $M_{\rm ex} \propto (\delta r)^{5/2}$ and $c_s^2 \propto \delta r$ one can show that the depth reached by the shock corresponds to an exterior mass of $M_{\rm ex}/M_* \simeq (p_{\rm ram}/p_\star)(t_{\rm cross}/t_{\rm sh})^{20/49}$
It is plausible to hypothesize that a fixed fraction of the shocked mass at this depth is stripped per collision, which would lead to $\Delta M/M_\star \propto (H_{\rm disk}/R_\star)^{20/49}$.   This dependence on $H_{\rm disk}$ is similar to, though somewhat stronger than what we find numerically in Figure \ref{fig:1st_hit} at small $H_{\rm disk}$.

The argument in this Appendix is only well-motivated when the star being shocked is an unperturbed $n = 3/2$ polytrope.  It thus only applies to the mass stripped in the `first' collision carried out in this paper, not the subsequent collisions.  The mass stripped per collision after many collisions has strongly modified the structure of the outer layers of the star is given by Equation \ref{eq:fit}.

\end{appendix}

\bibliography{sample631}{}
\bibliographystyle{aasjournal}

\end{CJK*}

\end{document}